\documentclass[a4paper]{article}

\usepackage{amsmath}
\usepackage{amssymb}
\usepackage{bbm}
\usepackage{graphicx}
\usepackage{natbib}

\newcommand{\diff}{\mathrm{d}}
\newcommand{\e}{\mathrm{e}}

\newcommand{\idty}{1\!\!1}

\newcommand{\R}{\text{I}\!\text{R}}

\newcommand{\tpl}[1]{\boldsymbol{#1}}

\newcommand{\Prob}{\text{Prob}}

\newcommand{\sS}{\mathcal{S}}
\newcommand{\sC}{\mathcal{C}}
\newcommand{\ds}{\displaystyle}
\newcommand{\ts}{\textstyle}
\newcommand{\dd}[2]{\frac{\partial{#1}}{\partial{#2}}}
\newcommand{\habi}{\mathcal{H}}

\newtheorem{result}{Result}
\newtheorem{assumption}{Assumption}
\newtheorem{definition}{Definition}

\begin{document}

\textbf{\LARGE Predicting Coexistence of Plants subject to \\[3pt] a Tolerance-Competition Trade-off}

\bigskip \bigskip

{\large Bart Haegeman, Tewfik Sari, Rampal S.~Etienne}

\bigskip \bigskip \bigskip

\section*{Abstract} 

Ecological trade-offs between species are often invoked to explain species coexistence in ecological communities.  However, few mathematical models have been proposed for which coexistence conditions can be characterized explicitly in terms of a trade-off.  Here we present a model of a plant community which allows such a characterization.  In the model plant species compete for sites where each site has a fixed stress condition.  Species differ both in stress tolerance and competitive ability.  Stress tolerance is quantified as the fraction of sites with stress conditions low enough to allow establishment.  Competitive ability is quantified as the propensity to win the competition for empty sites.  We derive the deterministic, discrete-time dynamical system for the species abundances.  We prove the conditions under which plant species can coexist in a stable equilibrium.  We show that the coexistence conditions can be characterized graphically, clearly illustrating the trade-off between stress tolerance and competitive ability.  We compare our model with a recently proposed, continuous-time dynamical system for a tolerance-fecundity trade-off in plant communities, and we show that this model is a special case of the continuous-time version of our model.

\bigskip \bigskip \bigskip \bigskip \bigskip \bigskip

{\small Bart Haegeman, Centre for Biodiversity Theory and Modelling, Experimental Ecology Station, Centre National de la Recherche Scientifique, Moulis, France.}

\bigskip

{\small Tewfik Sari, Irstea, UMR ITAP \& Modemic (Inra/Inria), UMR Mistea, Montpellier, France.}

\bigskip

{\small Rampal S. Etienne, Community and Conservation Ecology Group, Centre for Ecological and Evolutionary Studies, University of Groningen, Groningen, The Netherlands.}

\newpage
\section{Introduction} 

Coexistence of species in ecological communities is often explained by ecological trade-offs \citep{cl03,kc04}.  A trade-off expresses that the benefit of performing one ecological function well comes at the cost of performing another ecological function badly.  Well-known examples of trade-offs enabling local coexistence are the trade-off between efficiency of consumption of one resource and another \citep{m72,t82}, between the efficiency in consuming resources and the effectiveness of the defense to predators and parasites \citep{h94,u02,v10}, between competitive ability in resource consumption and tolerance of stressful abiotic factors \citep{tp93,cl03} and between competitive ability and the ability to colonize empty sites \citep{t82,r93,c06}.  These trade-offs allow temporal niche differention when the environment (i.e. resources, abiotic factors, predators) changes periodically.  Likewise, they allow regional coexistence when there is spatial variation in the environment.  This spatio-temporal variation opens up possibilities for coexistence of generalist and specialist strategies \citep{m96,wx12}.

While there is ample theoretical literature showing how trade-offs enable stable coexistence, there are relatively few simple predictive models of community assembly through trade-offs.  The description of trade-offs typically requires complex models with a large number of parameters which bars practical application.  Simple models with a limited number of parameters are essential for comparison of model predictions with empirical data.  In this paper we introduce such a simple model for the trade-off between competitive ability in resource consumption and tolerance of abiotic factors in a heterogeneous environment.

In the model plant species are characterized by two traits: tolerance of abiotic stress and competitive ability in resource consumption.  Stress tolerance is defined as the fraction of the environment in which a species can establish.  Species that are more stress tolerant are therefore more habitat generalists.  Competitive ability can be interpreted as the total propensity to colonize an empty site.  This interpretation is broader than just fecundity or total propagule biomass, because it allows differences in establishment that are not simply caused by mass effects.  Individuals of more competitive species are more likely to colonize empty sites than individuals of less competitive species.

After deriving the model equations, we analyze species coexistence at equilibrium.  Although the equilibrium and stability conditions are cumbersome to formulate analytically, we show that these conditions can be clearly represented graphically.  A graphical construction in the species trait plane allows us to easily predict which species coexist and in what relative abundances.  The main aim part of this paper is to establish the analytical conditions and justify their graphical interpretation.

Our model can be regarded as a generelization of the model proposed by \citet{ml10} for a tolerance-fecundity trade-off.  Our model is formulated in discrete time, but we show that the model of \citet{ml10} is a special case of the continuous-time version of our model. In particular, we show that the equilibrium and stability conditions that we prove for our discrete-time model, can be readily extended to the continuous-time dynamical system of \citet{ml10}.  This allows us to complement and correct \citet{ml10}'s partial analysis.

\section{Model equations} 
\label{sect:modeleqs}

In this section we construct the model equations of the ecological community model.  We consider here the case of sessile organisms (e.g., plants, corals, or territorial animals) with non-overlapping generations (i.e. annuals) leading to a discrete-time dynamical system.  We discuss the continuous-time dynamical system for overlapping generations in Section~\ref{sect:mlandau}.

\subsection*{General model}

\begin{figure}
\begin{center}
\includegraphics[width=\textwidth]{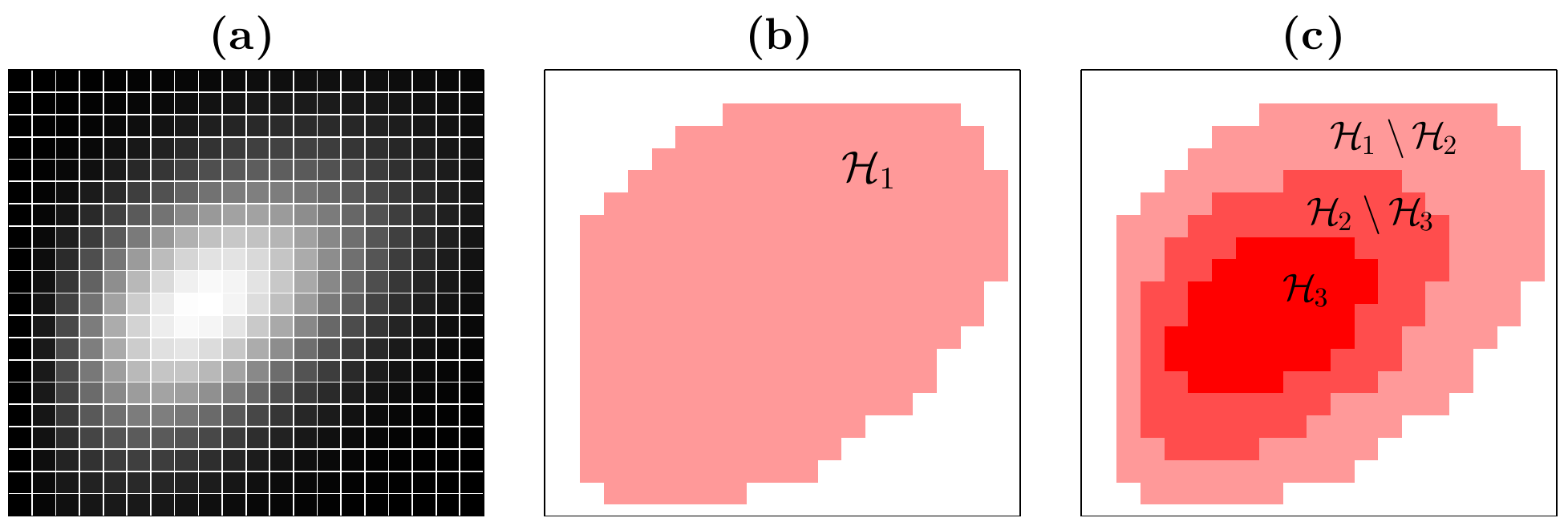}
\end{center}
\caption{Model accounts implicitly for spatial structure.  Panel~\textbf{a}: The environment consists of a large number of sites.  Each site is characterized by a local stress.  Darker colors denote more stressful conditions;  for the example shown here, the lowest stress is found in the center of the environment.  Panel~\textbf{b}: A species can only occupy sites in which the local stress is below the stress tolerance of the species.  The sites in which species 1 can establish, that is, the habitat $\habi_1$ of species 1, is shaded.  Panel~\textbf{c}: Less stress tolerant species have a smaller habitat.  Moreover, the habitat of a more stress tolerant species contains the habitat of a less stress tolerant species.  The sites in which different combinations of species 1, 2 and 3 can establish are shaded.  In habitat $\habi_1 \setminus \habi_2$ only species 1 can establish;  in habitat $\habi_2 \setminus \habi_3$ species 1 and 2 can establish, but not species 3;  in habitat $\habi_3$ all three species 1, 2 and 3 can establish.  The sizes of these habitats are $h_1-h_2$, $h_2-h_3$ and $h_3$, respectively.  Due to the model assumption that dispersal is uniform over the environment, the detailed spatial structure is irrelevant, that is, the model dynamics depend only on $h_1$, $h_2$ and $h_3$.} \label{fig:spatial}
\end{figure}

We consider an environment consisting of a large number of sites, see Figure~\ref{fig:spatial}.  Each site can be empty or occupied by one of $S$ species, depending on the species' stress tolerance:  a species can only establish in a site where the local stress is below its stress tolerance.  We denote the set of sites that species~$i$ can inhabit by $\habi_i$, the habitat of species $i$.  We denote the fraction of the environment that species~$i$ can inhabit by $h_i$.

The habitat size $h_i$ of species $i$ is a measure of its stress tolerance.  Larger habitat size $h_i$ implies a more stress tolerant species $i$.  We arrange the species in decreasing order of stress tolerance (we can do this without loss of generality), so that species~1 is the most stress tolerant, species~2 the second most stress tolerant, and so on.  We have
\begin{equation}
 \habi_S \subset \ldots \subset \habi_2 \subset \habi_1.
 \label{eq:habincl}
\end{equation}
If a specific species can establish in a specific site, then also all more stress tolerant species (that is, species with a smaller index) can establish in the site.  As a consequence,
\begin{equation}
 h_S < \ldots < h_2 < h_1.
 \label{eq:habineq}
\end{equation}

Denote by $p_i$ the fraction of the environment that is occupied by species $i$ at a certain time.  Because species can only occupy sites in its habitat $\habi_i$, we have $p_i \leq h_i$ for all $i$.  A stronger inequality holds because also species $i+1,i+2,\ldots,S$ can only occupy sites in habitat $\habi_i$, see (\ref{eq:habincl}).  Because a site cannot be occupied by more than one species, we have
\begin{equation}
 p_i + p_{i+1} + \ldots + p_S \leq h_i
 \qquad \text{for all $i$}.
\end{equation}

The species occupancies $p_1,p_2,\ldots,p_S$ change from one generation to the next one.  At the end of generation $t$ all individuals die and all sites of the environment are vacated.  At the beginning of generation $t+1$ propagules (e.g., seeds) produced and dispersed in generation $t$ can regenerate in the vacated sites.  As a result, the model applies particularly to annual plants.

To model the dispersal process, we assume that the number of propagules of a specific species to a specific site is a Poisson random variable.  These random variables are assumed to be mutually independent for different species and different sites.  All Poisson variables for species $i$ have the same parameter $f_i\,p_i$ with $f_i$ the fecundity of species $i$.  This implies that dispersal is uniform, that is, all sites receive the same propagule rain.

To model the recruitment process, we consider a site in $\habi_k \setminus \habi_{k+1}$, that is, a site which only species $1,2,\ldots,k$ (and no other species) can inhabit.  The propagules of species $k+1,\ldots,S$ cannot regenerate in this site, and can be neglected.  If no propagules of species $1,2,\ldots,k$ have dispersed to this site, then the site remains empty.  Otherwise, the site is occupied by one of the species $1,2,\ldots,k$.  To determine which species, one of the propagules that have dispersed to this site is drawn at random.

Due to these model assumptions, the spatial structure of the model is simplified considerably.  First, we have assumed that the dispersal process is uniform.  Therefore, we can discard the position of the sites occupied by different species;  it suffices to keep track of the species occupancies in different habitats $\habi_k \setminus \habi_{k+1}$.  Moreover, we have assumed that the fecundity $f_i$ of species $i$ is independent of the position of species $i$.  Hence, the number of propagules produced by species $i$ does not depend on the distribution of species $i$ over the habitats $\habi_k \setminus \habi_{k+1}$, but only on the species occupancy $p_i$ in the total environment.  As a result, the spatial structure of the model is taken into account implicitly, see Figure~\ref{fig:spatial}.

We construct the model equations for the dynamical variables $p_1,p_2,\ldots,p_S$.  First, we compute the probability $Q_{i|k}$ that species $i$ recruits a site in habitat $\habi_k \setminus \habi_{k+1}$.  A standard computation gives, see Appendix~A,
\begin{equation}
 Q_{i|k} = L\bigg(\sum_{j=1}^k f_j p_j\bigg) \, f_i p_i,
 \label{eq:probwin}
\end{equation}
with $L(x) = \frac{1-\e^{-x}}{x}$.  The probability $Q_{0|k}$ that a site in $\habi_k\setminus\habi_{k+1}$ remains empty is
\begin{equation}
 Q_{0|k} = \e^{-\sum_{i=1}^k f_i p_i}.
 \label{eq:probwin0}
\end{equation}
We interpret the probability $Q_{i|k}$ as the fraction of sites of habitat $\habi_k \setminus \habi_{k+1}$ recruited by species $i$.  This approximation is justified by the assumption that the number of sites is large.

The size of habitat $\habi_k \setminus \habi_{k+1}$ is equal to $h_k-h_{k+1}$.  Multiplying by (\ref{eq:probwin}), we get the size of the part of habitat $\habi_k \setminus \habi_{k+1}$ that is occupied by species $i$ in generation $t+1$.  Summing this product over $k \geq i$, we get the fraction of the environment that is occupied by species $i$ in generation $t+1$.  Hence, using the shorthand notation $p_i \equiv p_i(t)$ and $p_i^+ \equiv p_i(t+1)$,  we find
\begin{equation}
 p_i^+ = \sum_{k=i}^{S-1} (h_k-h_{k+1})\, Q_{i|k} + h_S\, Q_{i|S}.
 \label{eq:discgen}
\end{equation}
Equation~(\ref{eq:discgen}) for $i=1,2,\ldots,S$ defines the dynamical system for the species occupancies $p_1,p_2,\ldots,p_S$.

It is instructive to write down model~(\ref{eq:discgen}) for $S=2$ species,
\begin{subequations} \label{eq:disc2sp}
\begin{align}
 p_1^+ &= (h_1-h_2) \Big( 1-\e^{-f_1p_1} \Big) \nonumber \\
  &\quad + h_2 \frac{f_1p_1}{f_1p_1 + f_2p_2}
  \Big( 1-\e^{-\big(f_1p_1+f_2p_2\big)} \Big) \\[3pt]
 p_2^+ &= h_2 \frac{f_2p_2}{f_1p_1 + f_2p_2}
  \Big( 1-\e^{-\big(f_1p_1+f_2p_2\big)} \Big).
\end{align}
\end{subequations}
Recall that species 1 is more stress tolerant than species 2 (because $h_1>h_2$, see (\ref{eq:habineq})).  Hence, whereas species 2 can only recruit in habitat $\habi_2$, species 1 can recruit in habitats $\habi_1\setminus\habi_2$ and $\habi_2$, corresponding to the two terms in (\ref{eq:disc2sp}a).  In habitat $\habi_1\setminus\habi_2$ species 1 is the only species to recruit sites;  a fraction $\e^{-f_1p_1}$ of these sites remains empty.  In habitat $\habi_2$ both species 1 and 2 can recruit sites;  a fraction $\e^{-(f_1p_1+f_2p_2)}$ of these sites remains empty.  The occupied sites are attributed to species 1 and 2 proportional to $f_1p_1$ and $f_2p_2$, respectively.

\subsection*{Limiting model for large fecundities}

By simultaneously increasing the fecundities $f_i$, the number of empty sites diminishes.  This suggests a limiting case of the general model~(\ref{eq:discgen}) in which all sites are constantly occupied.  To describe this limiting case, we introduce the fecundity scaling factor $\alpha$ and rescaled fecundities $\widehat f_i$.  We consider the coupled limit
\begin{equation}
 \alpha\to\infty \quad\text{and}\quad f_i\to\infty \quad
 \text{with}\ f_i = \alpha \widehat f_i \quad \text{for all $i$}.
 \label{eq:scaling}
\end{equation}
In the limit~(\ref{eq:scaling}) probability~(\ref{eq:probwin}) of species $i$ recruiting an empty site in habitat $\habi_k\setminus\habi_{k+1}$ is
\begin{equation}
 \widehat Q_{i|k} = 
 \begin{cases}
 \ds \frac{\widehat f_i p_i}{\ts\sum_{j=1}^k \widehat f_j p_j}
 & \text{if $p_i>0$} \\[12pt]
 0 & \text{if $p_i=0$.}
 \end{cases}
 \label{eq:defhatQ}
\end{equation}
Hence, the site is occupied with certainty, except if all species that can recruit the site have become extinct.  Species $i$ occupies the site with a probability proportional to $\widehat f_i p_i$.  The corresponding dynamical system is
\begin{equation}
 p_i^+ = \sum_{k=i}^{S-1} (h_k-h_{k+1})\, \widehat Q_{i|k} + h_S\, \widehat Q_{i|S}.
 \label{eq:disclim}
\end{equation}
The limiting model~(\ref{eq:disclim}) is important because in the following sections we formulate some of the results for the general model~(\ref{eq:discgen}) in terms of the limiting model~(\ref{eq:disclim}).

\section{Conditions for coexistence at equilibrium} 
\label{sect:condcoexist}

Dynamical system~(\ref{eq:discgen}) describes a community of $S$ species.  Each species $i$ is characterized by two traits, its stress tolerance $h_i$ and its fecundity $f_i$.  In this section we give the conditions on the species traits $h_i$ and $f_i$ for species coexistence at equilibrium.  The (local) stability of the equilibria is studied in the next section.

We denote the set of all $S$ species by $\sS = \{1,2,\ldots,S\}$.  Each subset $\sC\subset\sS$ corresponds to a potential coexistence equilibrium.

\begin{definition}[Coexistence equilibrium]
Consider a subset $\sC \subset \sS$.  We call an equilibrium $(\widetilde p_1,\widetilde p_2,\ldots, \widetilde p_S)$ of dynamical system~(\ref{eq:discgen}) an equilibrium with coexistence set $\sC$ if $\widetilde p_i>0$ for all $i \in \sC$ and $\widetilde p_i=0$ for all $i \in \sS\setminus\sC$.
\end{definition}

To formulate the coexistence conditions, it is convenient to introduce the inverse fecundities $g_i = \frac{1}{f_i}$.  We exclude non-generic parameter combinations by the following assumption.

\begin{assumption}[Genericity of parameters] \label{ass:generic}
We assume that the species traits $h_i$ and $f_i$, $i=1,2,\ldots,S$, satisfy the following conditions: \\[-18pt]
\begin{itemize}
\item the parameters $h_i$, $i=1,2,\ldots,S$, are positive and arranged in decreasing order,
\begin{equation}
 h_1 > h_2 > \ldots > h_S;
 \label{eq:habineqq}
\end{equation}
\item the parameters $f_i$, $i=1,2,\ldots,S$, are positive and mutually different;
\item the following ratios are mutually different and different from 1:
\[
 \frac{h_i}{g_i}, \ i=1,2,\ldots,S
 \qquad \text{and} \qquad
 \frac{h_i-h_j}{g_i-g_j}, \ i,j=1,2,\ldots,S, \ i\neq j,
\]
with $g_i = \frac{1}{f_i}$.
\end{itemize}
\end{assumption}

The following result is proved in Appendix~B.

\begin{result}[Existence of equilibrium] \label{res:equilib}
Suppose Assumption~\ref{ass:generic} is verified.  Consider a subset $\sC = \{i_1,i_2,\ldots,i_C\} \subset \sS$ with $i_1<i_2<\ldots<i_C$.  Then there exists an equilibrium with coexistence set $\sC$ if and only if the following inequalities are satisfied:
\begin{equation}
 1 < \frac{h_{i_1}-h_{i_2}}{g_{i_1}-g_{i_2}}
   < \frac{h_{i_2}-h_{i_3}}{g_{i_2}-g_{i_3}}
   < \ldots
   < \frac{h_{i_{C-1}}-h_{i_C}}{g_{i_{C-1}}-g_{i_C}}
   < \frac{h_{i_C}}{g_{i_C}}.
 \label{eq:condexistsome}
\end{equation}
If an equilibrium $(\widetilde p_1,\widetilde p_2,\ldots, \widetilde p_S)$ with coexistence set $\sC$ exists, then the equilibrium occupancies $\widetilde p_{i_k} > 0$ are given by 
\begin{subequations} \label{eq:equilib}
\begin{align}
 \widetilde p_{i_1} &=
 g_{i_1} \; \Lambda\bigg(\frac{h_{i_{1}}-h_{i_{2}}}{g_{i_{1}}-g_{i_{2}}}\bigg) \\
 \widetilde p_{i_k} &=
 g_{i_k} \Bigg( \Lambda\bigg(\frac{h_{i_{k}}-h_{i_{k+1}}}{g_{i_{k}}-g_{i_{k+1}}}\bigg)
 - \Lambda\bigg(\frac{h_{i_{k-1}}-h_{i_{k}}}{g_{i_{k-1}}-g_{i_{k}}}\bigg) \Bigg)
 \qquad \text{$k = 2,3,\ldots,C-1$} \\
 \widetilde p_{i_C} &=
 g_{i_C} \Bigg( \Lambda\bigg(\frac{h_{i_{C}}}{g_{i_{C}}}\bigg)
 - \Lambda\bigg(\frac{h_{i_{C-1}}-h_{i_{C}}}{g_{i_{C-1}}-g_{i_{C}}}\bigg) \Bigg),
\end{align}
\end{subequations}
where $\Lambda$ denotes the inverse function of $x \mapsto \frac{x}{1-\e^{-x}}$.  Finally, if an equilibrium with coexistence set $\sC$ exists, then there is no other equilibrium with coexistence set $\sC$.
\end{result}

Note that the function $\Lambda$ in Result~\ref{res:equilib} can be expressed in terms of the upper branch $W_0$ of the Lambert $W$ function,
\begin{equation*}
 \Lambda(x) = x + W_0(-x\,\e^{-x}),
 \label{eq:lamb0}
\end{equation*}
see Appendix~B for a derivation.

From (\ref{eq:habineqq}) and (\ref{eq:condexistsome}) it follows that
\begin{equation}
 g_{i_1} > g_{i_2} > \ldots > g_{i_{C-1}} > g_{i_C},
 \label{eq:fecineq}
\end{equation}
or
\[
 f_{i_1} < f_{i_2} < \ldots < f_{i_{C-1}} < f_{i_C}.
\]
Species can coexist only if less tolerant species are more fecund, which indicates a trade-off between stress tolerance and fecundity.  However, condition~(\ref{eq:fecineq}) is necessary but not sufficient for species coexistence.  The tolerance-competition trade-off imposes stonger conditions than (\ref{eq:fecineq}).

\begin{figure}
\begin{center}
\includegraphics[width=.8\textwidth]{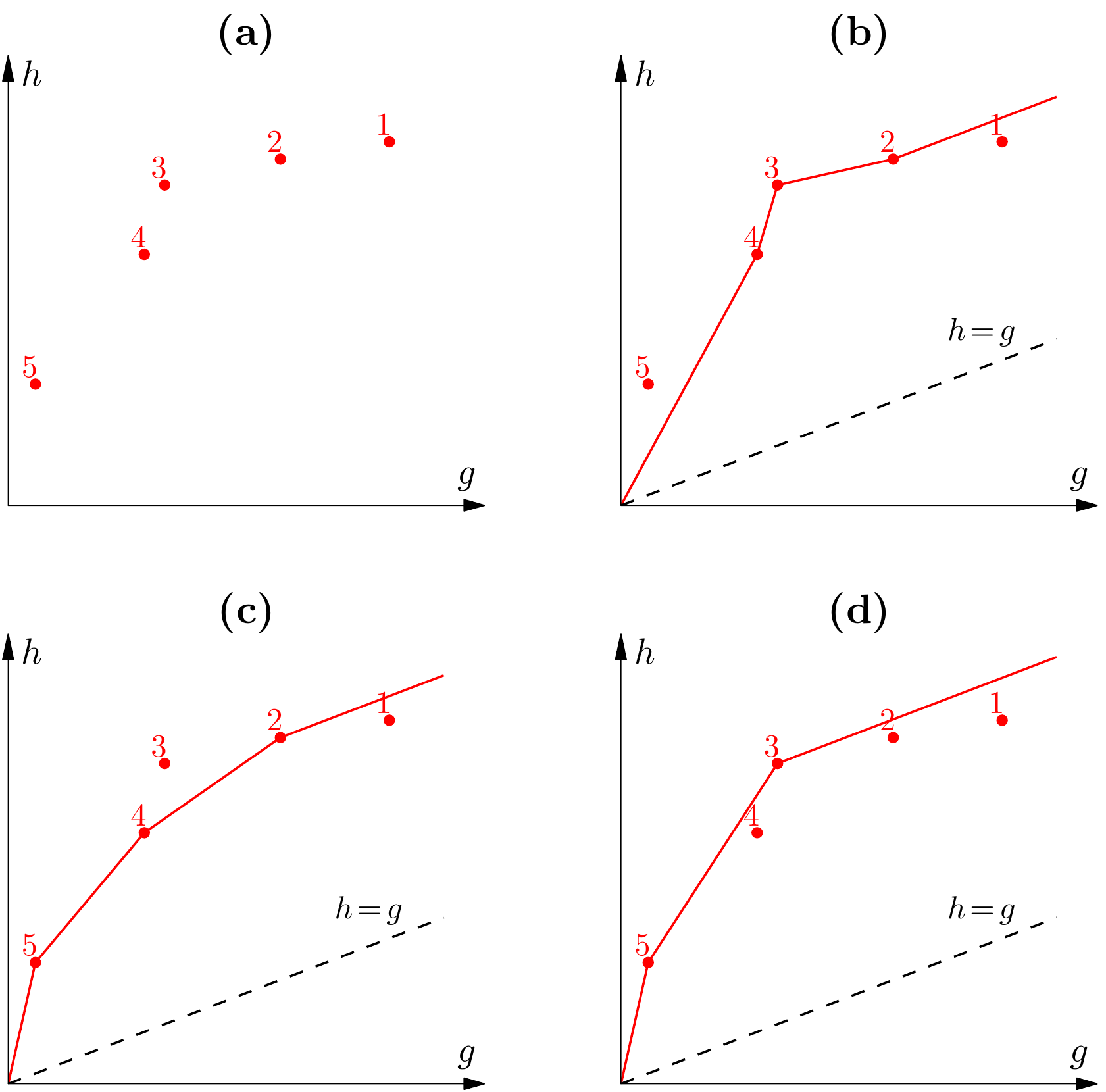}
\end{center}
\caption{Graphical representation of the coexistence conditions~(\ref{eq:condexistsome}) and the stability conditions~(\ref{eq:condstable}).  Panel~\textbf{a}:  The traits of $S=5$ species are represented in the $(g,h)$ plane; $g_i=\frac{1}{f_i}$ are inverse fecundities and $h_i$ are stress tolerances.  Panel~\textbf{b}:  We check whether an equilibrium exists in which species 2, 3 and 4 coexist, and species 1 and 5 are absent (corresponding to $\sC = \{2,3,4\}$).  We draw a broken line through the species traits $(g_i,h_i)$ with $i\in\sC$; the leftmost line segment passes through the origin; the rightmost line segment is parallel to the $h=g$ line (dashed line).  Because the resulting function is not concave, there does not exist such an equilibrium.  Panel~\textbf{c}:  We check whether an equilibrium exists in which species 2, 4 and 5 coexist, and species 1 and 3 are absent ($\sC = \{2,4,5\}$).  The same construction as in panel~\textbf{b} gives a concave broken line.  Hence, there exists such an equilibrium.  Because the traits of species 3 lie to the left of and above the broken line, the equilibrium is unstable.  Panel~\textbf{d}:  An equilibrium exists in which species 3 and 5 coexist, and species 1, 2 and 4 are absent ($\sC = \{3,5\}$).  Because the traits of all absent species lie to the right of and below the broken line, the equilibrium is stable.}
\label{fig:condexist}
\end{figure}

The necessary and sufficient condition for species coexistence can be represented graphically, see Figure~\ref{fig:condexist}.  We represent the traits $(g_i,h_i)$ of all species in the $(g,h)$ plane (Figure~\ref{fig:condexist}\textbf{a}).  To verify whether an equilibrium exists with coexistence set $\sC$, we draw a broken line through the points corresponding to the species in set $\sC$ (Figure~\ref{fig:condexist}\textbf{b}--\textbf{d}).  The broken line consists of the following line segments: \\[-18pt]
\begin{itemize}
\item a line segment between species $i_k$ and $i_{k+1}$, for $k=1,2,\ldots,C-1$;
\item a line segment from species $i_1$ with slope one towards larger $g$;
\item a line segment from species $i_C$ to the origin $(g=0,h=0)$.
\end{itemize}
It follows from inequalities~(\ref{eq:habineqq}) and (\ref{eq:fecineq}) that if there is an equilibrium with coexistence set $\sC$, then the broken line should define an increasing function of $g$.  Condition~(\ref{eq:condexistsome}) expresses that the slopes of the line segments should decrease for increasing $g$.  This corresponds to requiring that the function defined by the broken line is concave.  If this condition is fulfilled, then there exists an equilibrium with coexistence set $\sC$.

Examples of the broken-line construction are shown in Figure~\ref{fig:condexist}\textbf{b}--\textbf{d}.  In panel~\textbf{b} we check the coexistence of $\sC=\{2,3,4\}$.  The broken line is not concave, so that no equilibrium exists with coexistence set $\sC$.  In panels~\textbf{c} and \textbf{d} we check the coexistence of $\sC=\{2,4,5\}$ and $\sC=\{3,5\}$, respectively.  In both cases the broken line is concave, so that an equilibrium exists with coexistence set $\sC$ coexist.  For a specific set $\sC$ of coexisting species, the equilibrium, if it exists, is unique.  However, there can be different sets $\sC$ of coexisting species for which an equilibrium exists.  We show in Section~\ref{sect:stabcoexist} that only one of these equilibria is stable.

\begin{figure}
\begin{center}
\includegraphics[width=.95\textwidth]{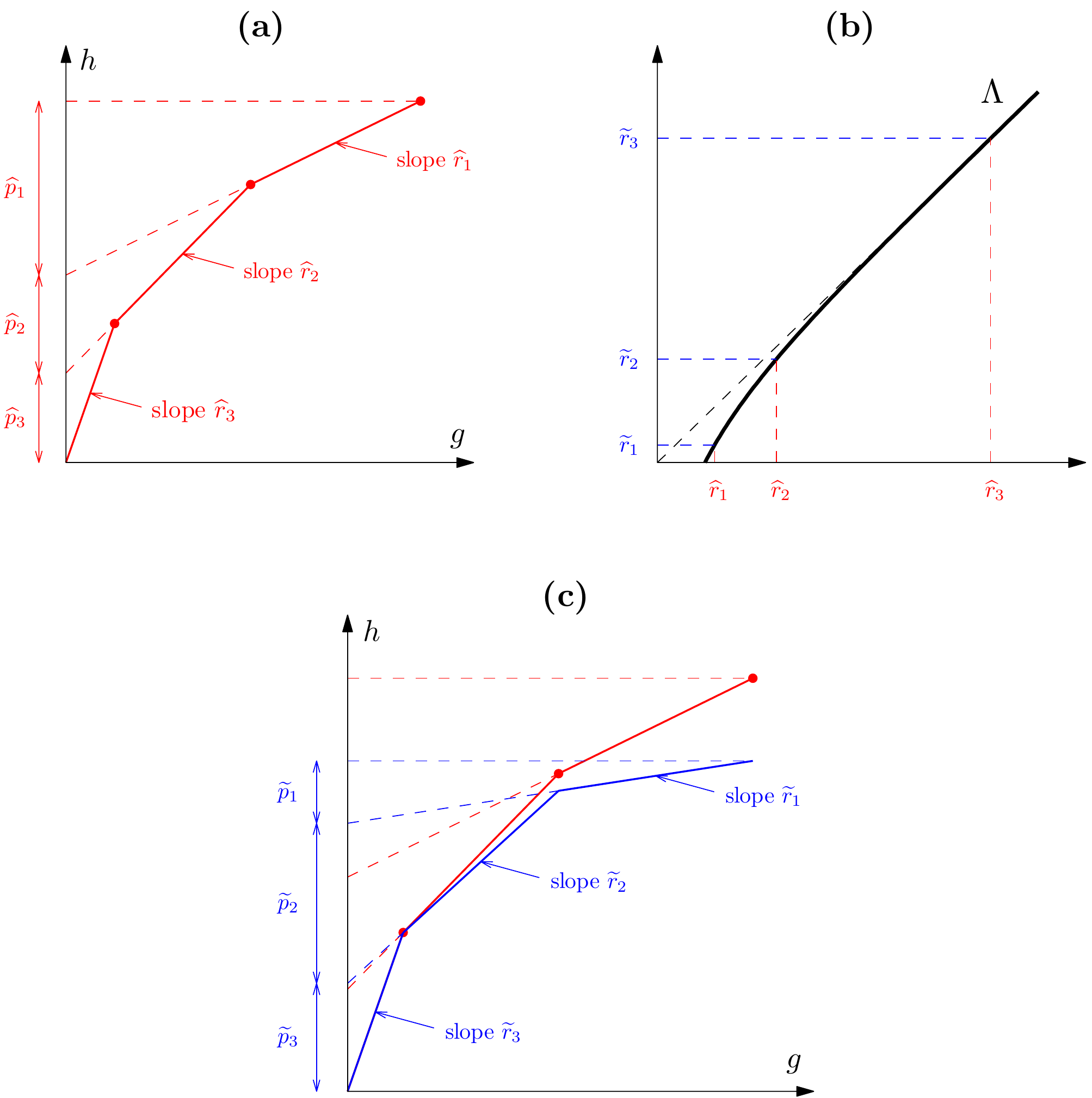}
\end{center}
\caption{Graphical construction of the equilibrium occupancies.  Panel~\textbf{a}: We draw the traits of the species in the $(g,h)$ plane (red dots; there are $S=3$ species), and construct a broken line through the species traits, as shown in the figure (solid red line).  By extending the line segment towards the $g=0$ axis (dashed red line), we obtain the equilibrium occupancies $\widehat p_i$.  Panel~\textbf{b}:  For each of the $S$ slopes of the broken line in panel~\textbf{a}, we compute a transformed slope using the function $\Lambda$.  Panel~\textbf{c}:  We construct a broken line with the transformed slopes, as shown in the figure.  Because the transformed slopes $\widetilde r_i = \Lambda(\widehat r_i)$ are smaller than the initial slopes $\widehat r_i$, the new broken line (solid blue line) is a squeezed version of the initial broken line (solid red line).  Using the same construction as in panel~\textbf{a}, that is, by extending the line segments towards the $g=0$ axis (dashed blue line), we obtain the equilibrium occupancies $\widetilde p_i$.}
\label{fig:equilib}
\end{figure}

A graphical construction for the equilibrium occupancies $\widetilde p_{i_k}$ is given in Figure~\ref{fig:equilib}.  We start with the limiting model~(\ref{eq:disclim}) for which Result~\ref{res:equilib} also holds.  The limit (\ref{eq:scaling}) corresponds to replacing in (\ref{eq:equilib}) the function $\Lambda$ with the identity function.  The resulting equilibrium occupancies $\widehat p_{i_k}$ are
\begin{subequations} \label{eq:withoutseedlim}
\begin{align}
 \widehat p_{i_1} &=
 g_{i_1} \; \frac{h_{i_{1}}-h_{i_{2}}}{g_{i_{1}}-g_{i_{2}}} \\
 \widehat p_{i_k} &=
 g_{i_k} \bigg( \frac{h_{i_{k}}-h_{i_{k+1}}}{g_{i_{k}}-g_{i_{k+1}}}
 - \frac{h_{i_{k-1}}-h_{i_{k}}}{g_{i_{k-1}}-g_{i_{k}}} \bigg)
 \qquad \text{$k = 2,3,\ldots,C-1$} \\
 \widehat p_{i_C} &=
 g_{i_C} \bigg( \frac{h_{i_{C}}}{g_{i_{C}}}
 - \frac{h_{i_{C-1}}-h_{i_{C}}}{g_{i_{C-1}}-g_{i_{tC}}} \bigg).
\end{align}
\end{subequations}
The broken-line representation introduced in Figure~\ref{fig:condexist} can be used to construct the occupancies $\widehat p_{i_k}$, see Figure~\ref{fig:equilib}\textbf{a}.  The difference between (\ref{eq:equilib}) and (\ref{eq:withoutseedlim}) resides in the function $\Lambda$.  We use the function $\Lambda$ to transform the slopes of the broken line of Figure~\ref{fig:equilib}\textbf{a}, see Figure~\ref{fig:equilib}\textbf{b}.  We construct a new broken line with the transformed slopes, from which we obtain the occupancies $\widetilde p_{i_k}$, see Figure~\ref{fig:equilib}\textbf{c}.

Finally, we note that the last statement of Result~\ref{res:equilib} says that any two equilibria with the same coexistence set are the same equilibrium.  Similar results have been reported for other dynamical systems, e.g., Boolean networks \citep{vxx}.

\section{Stability of coexistence equilibrium} 
\label{sect:stabcoexist}

In Result~\ref{res:equilib} we have characterized the equilibria of dynamical system~(\ref{eq:discgen}).  There can be up to $2^S$ equilibria for a community of $S$ species.  In this section we study the local stability of the equilibria.  We show that there is only one locally exponentially stable equilibrium.

We state the stability conditions for an equilibrium of dynamical system~(\ref{eq:discgen}).  The result is proved in Appendix~C.

\begin{result}[Stability of equilibrium] \label{res:stable}
Suppose Assumption~\ref{ass:generic} is verified.  Consider a subset $\sC = \{i_1,i_2,\ldots,i_C\} \subset \sS$ with $i_1<i_2<\ldots<i_C$.  Consider an equilibrium $(\widetilde p_1,\widetilde p_2,\ldots, \widetilde p_S)$ of dynamical system~(\ref{eq:discgen}) with coexistence set $\sC$.  Then the equilibrium is locally exponentially stable if and only if the following inequalities are satisfied:
\begin{subequations} \label{eq:condstable}
\begin{align}
 1 > \frac{h_{j}-h_{i_1}}{g_{j}-g_{i_1}}
 & \qquad \text{for all $j\in\sS\setminus\sC$ for which $j<i_1$} \\
 \frac{h_{i_k}-h_{j}}{g_{i_k}-g_{j}} > \frac{h_{j}-h_{i_{k+1}}}{g_{j}-g_{i_{k+1}}}
 & \qquad \text{for all $j\in\sS\setminus\sC$ and $k\in\{1,2,\ldots,C-1\}$} \nonumber \\[-8pt]
 & \qquad \text{for which $i_{k}<j<i_{k+1}$} \\  
 \frac{h_{i_C}-h_{j}}{g_{i_C}-g_{j}} > \frac{h_{j}}{g_{j}}
 & \qquad \text{for all $j\in\sS\setminus\sC$ for which $i_{C}<j$}.
\end{align}
\end{subequations}
\end{result}

We note that Result~\ref{res:stable} implies that the equilibrium is locally asymptotically stable.  Also note that the equilibrium cannot be asymptotically stable without being exponentially stable because Assumption~\ref{ass:generic} excludes the case in which some eigenvalues have modulus one.

Conditions~(\ref{eq:condstable}) express that each of the $S-C$ species absent at equilibrium cannot invade the community of the $C$ coexisting species.  Non-invasibility is a general necessary condition for stable species coexistence.  Result~\ref{res:stable} says that for dynamical system~(\ref{eq:discgen}) non-invasibility is also a sufficient condition.

The stability conditions~(\ref{eq:condstable}) can be formulated equivalently as
\begin{subequations} \label{eq:condstablealt}
\begin{align}
 h_j - h_{i_1} &< g_j - g_{i_1} \\[2pt]
 h_j - h_{i_{k+1}} &< \frac{h_{i_k} - h_{i_{k+1}}}
 {g_{i_k} - g_{i_{k+1}}}\, (g_j - g_{i_{k+1}})
 \qquad \text{for all $k\in\{1,2,\ldots,C-1\}$} \\
 h_j &< \frac{h_{i_C}}{g_{i_C}}\, g_j,
\end{align}
\end{subequations}
for all $j \in \sS\setminus\sC$.  Conditions~(\ref{eq:condstablealt}) can be interpreted graphically, see Figure~\ref{fig:condexist}.  We have explained how to determine whether a specific set $\sC$ of $C$ species can coexist at equilibrium.  The construction is based on a broken line containing the trait pairs $(g_i,h_i)$ of the $C$ coexisting species.  Conditions~(\ref{eq:condstablealt}) impose that the trait pairs $(g_i,h_i)$ of the other $S-C$ species absent at equilibrium should lie to the right of and below this broken line.  For example, in panel~\textbf{c} we check the stability of the coexistence equilibrium for $\sC = \{2,4,5\}$.  Species 3 lies to the left of and above the broken line, so the equilibrium is unstable.  In panel~\textbf{d} we check the stability of the coexistence equilibrium for $\sC = \{3,5\}$.  All species lie to the right of and below the broken line, so the equilibrium is locally stable.

In Figure~\ref{fig:condexist} we have considered a community of $S=5$ species.  It can be checked graphically that the broken line corresponding to $\sC = \{3,5\}$ is the only broken line satisfying the conditions imposed by inequalities~(\ref{eq:condexistsome}) and (\ref{eq:condstable}).  That is, there is only one locally stable equilibrium.    This suggests the following general result, proved in Appendix~D:

\begin{result}[Uniqueness of stable equilibrium] \label{res:onestable}
Suppose Assumption~\ref{ass:generic} is verified.  Then there is one and only one locally asymptotically stable equilibrium of dynamical system~(\ref{eq:discgen}).
\end{result}

\section{Continuous-time model} 
\label{sect:mlandau}

In this section we compare our model with the model proposed by \citet{ml10}.  Contrary to discrete-time dynamical system (\ref{eq:discgen}), the model of \citet{ml10} is formulated in continuous time.  To allow comparison we construct the continuous-time equivalent of model~(\ref{eq:discgen}), and argue that the model of \citet{ml10} can be recovered as a limiting model.  We show that the equilibrium and stability results for model~(\ref{eq:discgen}) also apply to the model of \citet{ml10}.

The discrete-time model (\ref{eq:discgen}) assumes that the dynamics between two generations consist of two processes: first, all individuals die, vacating the occupied sites, and second, some of the sites are recruited by a new individual.  In the continuous-time model, these two processes are no longer separated.  Individuals die and vacate their site in a continuous manner, and simultaneously, sites are recruited by a new individual in a continuous manner.

We introduce the mortality rate $m$ and the recruitment rate $r$, both of which are assumed to be site- and species-independent.  The recruitment process is modelled in the same way as in the discrete-time model.  In an empty site of habitat $\habi_k\setminus\habi_{k+1}$, propagules of species $1,2,\ldots,k$ can regenerate.  The probability that species $i$ recruits the site is given by $Q_{i|k}$ in (\ref{eq:probwin}).  Note that there is a non-zero probability $Q_{0|k}$ that the site remains empty (until a following recruitment event).  Hence, the effective recruitment rate for an empty site in $\habi_k\setminus\habi_{k+1}$ is $(1-Q_{0|k}) r \leq r$.

We aim to construct the model equations for the dynamical variables $p_1,p_2,\ldots,p_S$.  The rate of decrease of occupancy $p_i$ due to mortality is $m\,p_i$.  The rate of increase of occupancy $p_i$ due to recruitment in habitat $\habi_k\setminus\habi_{k+1}$ is equal to $r\,Q_{i|k}$ times the size of the set of empty sites in $\habi_k\setminus\habi_{k+1}$.  However, we cannot express the latter factor in terms of the species occupancies $p_1,p_2,\ldots,p_S$.  We know that \\[-6pt]
\[
\text{size of $\big\{$empty sites in habitat $\habi_1$$\big\}$} = h_1 - \sum_{i=1}^S p_i, \\[2pt]
\]
but we do not know the distribution of the empty sites over the habitats $\habi_k\setminus\habi_{k+1}$.  The variables $p_1,p_2,\ldots,p_S$ are not sufficiently informative to construct the model equations.

To solve the problem, we extend the set of dynamical variables.  We introduce the set of occupancies $p_{ik}$ defined by
\[
 p_{ik} = \text{size of $\big\{$sites occupied by species $i$ in habitat $\habi_k\setminus\habi_{k+1}$$\big\}$}.
\]
We have \\[-20pt]
\begin{align*}
 & \text{size of $\big\{$sites occupied by species $i$$\big\}$}
  = \sum_{k=i}^S p_{ik} = p_i \\[-6pt]
 & \text{size of $\big\{$occupied sites in habitat $\habi_k\setminus\habi_{k+1}$$\big\}$}
  = \sum_{i=1}^k p_{ik} \\[-6pt]
 & \text{size of $\big\{$empty sites in habitat $\habi_k\setminus\habi_{k+1}$$\big\}$}
  = h_k-h_{k+1} - \sum_{i=1}^k p_{ik}. \\[-18pt]
\end{align*}
The rate of decreases of occupancy $p_{ik}$ due to mortality is $m\,p_{ik}$.  The rate of increase of occupancy $p_{ik}$ due to recruitment is equal to $r\,Q_{i|k}$ times the size of the set of empty sites in habitat $\habi_k\setminus\habi_{k+1}$.  Hence,
\begin{equation}
 \frac{\diff p_{ik}}{\diff t} = - m\,p_{ik} +
 r \bigg( h_k-h_{k+1} - \sum_{j=1}^k p_{jk} \bigg)\,Q_{i|k}.
 \label{eq:contfull}
\end{equation}
Recall that the probabilities $Q_{i|k}$ depend on the occupancies $p_i$, which can be expressed in terms of the occupancies $p_{ik}$.  Hence, equation (\ref{eq:contfull}) for $i=1,2,\ldots,S$ and $k=i,i+1,\ldots,S$ defines an autonomous dynamical system.

It is interesting to note that the above-mentioned problem is not present in the limiting case of large recruitment rate $r$.  If the recruitment rate $r$ is large, all sites are constantly occupied;  as soon as an individual dies, a new individual recruits the vacated site.  In that case, the effective rate of recruitment in $\habi_k\setminus\habi_{k+1}$ equals the rate of deaths in $\habi_k\setminus\habi_{k+1}$, given by $m (h_k-h_{k+1})$.  The model equations are then
\begin{align}‌
 \frac{\diff p_i}{\diff t}
 &= - m\,p_i + m \bigg( \sum_{k=i}^{S-1}
  (h_k-h_{k+1})\,\frac{Q_{i|k}}{1-Q_{0|k}}
 + h_S\,\frac{Q_{i|S}}{1-Q_{0|S}} \bigg) \nonumber \\
 &= - m\,p_i + m \bigg( \sum_{k=i}^{S-1}
  (h_k-h_{k+1})\,\widehat Q_{i|k}
 + h_S\,\widehat Q_{i|S} \bigg). \label{eq:contlim}
\end{align}
with the probability $\widehat Q_{i|k}$ given in (\ref{eq:defhatQ}).  Dynamical system (\ref{eq:contlim}) is identical to the model ``without seed limitation'' of \citet{ml10}.  Because in the derivation we have assumed that all sites are occupied, dynamical system~(\ref{eq:contlim}) is restricted to the hyperplance $\sum_{i=1}^S p_i = h_1$.  It is easily verified that this hyperplane is invariant under (\ref{eq:contlim}).

\citet{ml10} also introduced a model ``with seed limitation'' defined by
\begin{equation}
 \frac{\diff p_i}{\diff t}
  = - m\,p_i + m \bigg( \sum_{k=i}^{S-1}
  (h_k-h_{k+1})\,Q_{i|k} + h_S\,Q_{i|S} \bigg).
 \label{eq:contgen}
\end{equation}
Dynamical system (\ref{eq:contgen}) is obtained from dynamical system (\ref{eq:contlim}) by replacing the limiting probabilities $\widehat Q_{i|k}$ for large fecundities (that is, without seed limitation) by the probabilities $Q_{i|k}$ for finite fecundities (that is, with seed limitation).  However, equation (\ref{eq:contgen}) seems to be missing a consistent interpretation.  In particular, it is unclear how the problem of the insufficiency of the variables $p_1,p_2,\ldots,p_S$ is solved.  The workaround of model (\ref{eq:contlim}) does not apply, because model (\ref{eq:contgen}) allows empty sites (due to seed limitation).  Indeed, the hyperplane $\sum_{i=1}^S p_i = h_1$ is not invariant under dynamical system (\ref{eq:contgen}).  For these reasons, we suggest that the model of \citet{ml10} with seed limitation should be replaced by the higher-dimensional but consistent model~(\ref{eq:contfull}).

\begin{figure}
\begin{center}
\includegraphics[width=.9\textwidth]{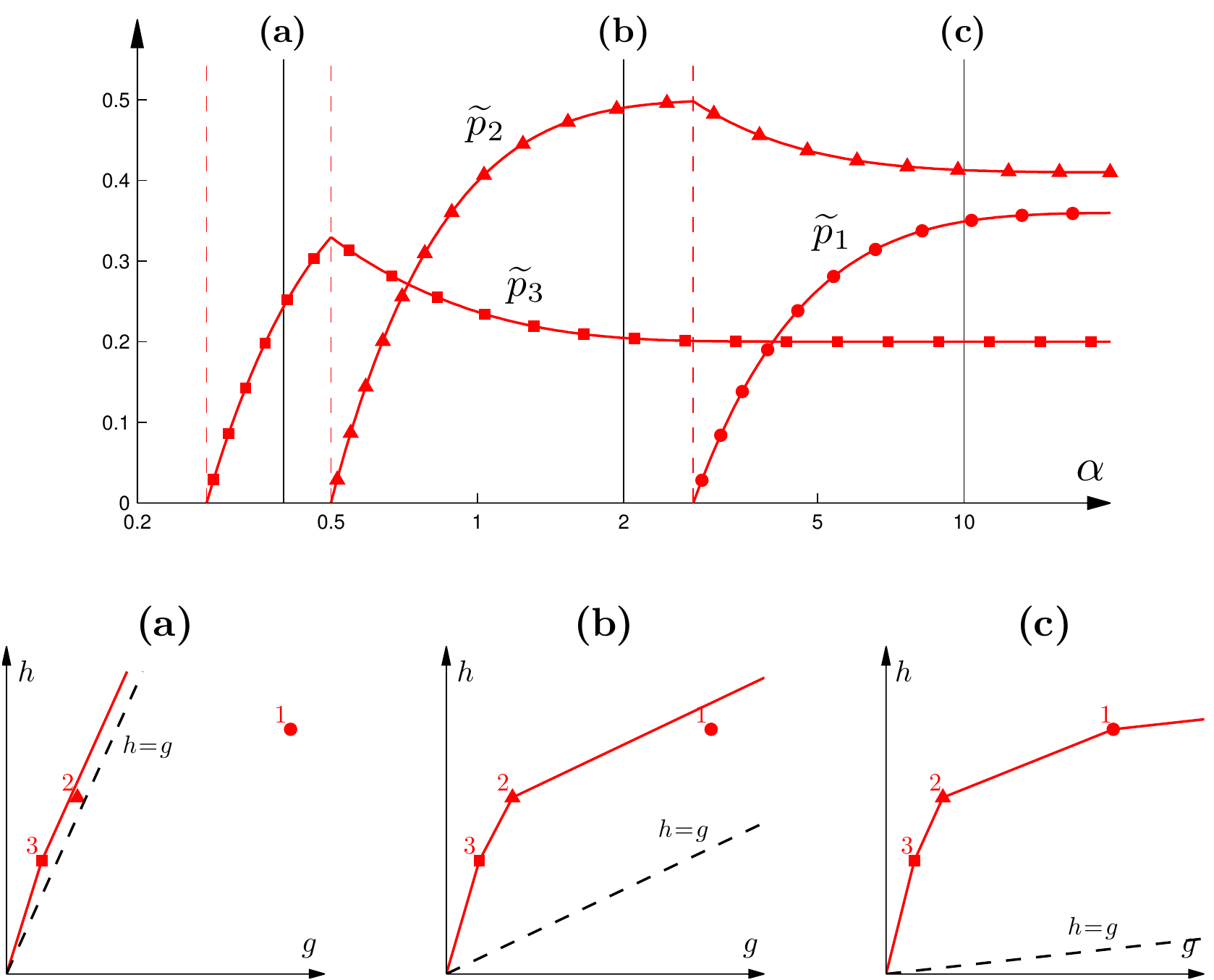}
\end{center}
\caption{Equilibrium community composition for increasing species fecundities.  We consider a community consisting of $S=3$ species.  For each species $i$ the fecundity $f_i$ is proportional to a fecundity scaling factor $\alpha$: $f_1=\alpha$ and $h_1=0.97$ for species 1; $f_2=4\alpha$ and $h_2=0.7$ for species 2; $f_3=8\alpha$ and $h_3=0.45$ for species 3.  Main panel:  Equilibrium species occupancies $\widetilde p_1,\;\widetilde p_2,\;\widetilde p_3$ are plotted as a function of fecundity scaling factor $\alpha$.  Circle markers: species 1; triangle markers: species 2; square markers: species 3.  We distinguish four parameter zones (separated by vertical dashed lines):  for $\alpha\leq\frac{5}{18}$ none of the species can survive;  for $\frac{5}{18}<\alpha\leq 0.5$ only the most fecund species survives;  for $0.5<\alpha\leq\frac{25}{9}$ the two most fecund species coexist;  for $\alpha>\frac{25}{9}$ all three species coexist.  Subpanels~\textbf{a}, \textbf{b} and \textbf{c} show the community composition for $\alpha=0.4$, $\alpha=2$ and $\alpha=10$, respectively.  The scale of the $g$-axis differs between subpanels.}
\label{fig:seedlim}
\end{figure}

The lack of a clear interpretation of dynamical system (\ref{eq:contgen}) does not prohibit its mathematical analysis.  Interestingly, our analysis of model~(\ref{eq:discgen}) can be extended to model~(\ref{eq:contgen}).  We prove in Appendix~E that model~(\ref{eq:contgen}) has the same equilibrium and stability conditions as model~(\ref{eq:discgen}).  This result complements and corrects a number of results reported in \citet{ml10}, see Appendix~E.  To illustrate the correspondence between both models, we analyze an example system in Figure~\ref{fig:seedlim}.  The curves were computed using the analytical results for model~(\ref{eq:discgen}).  The figure agrees well with Figure~3C in \citet{ml10}, which was generated by simulating equations~(\ref{eq:contgen}) for the same parameter values.

\section{Discussion} 
\label{sect:discuss}

\begin{figure}
\begin{center}
\includegraphics[width=.8\textwidth]{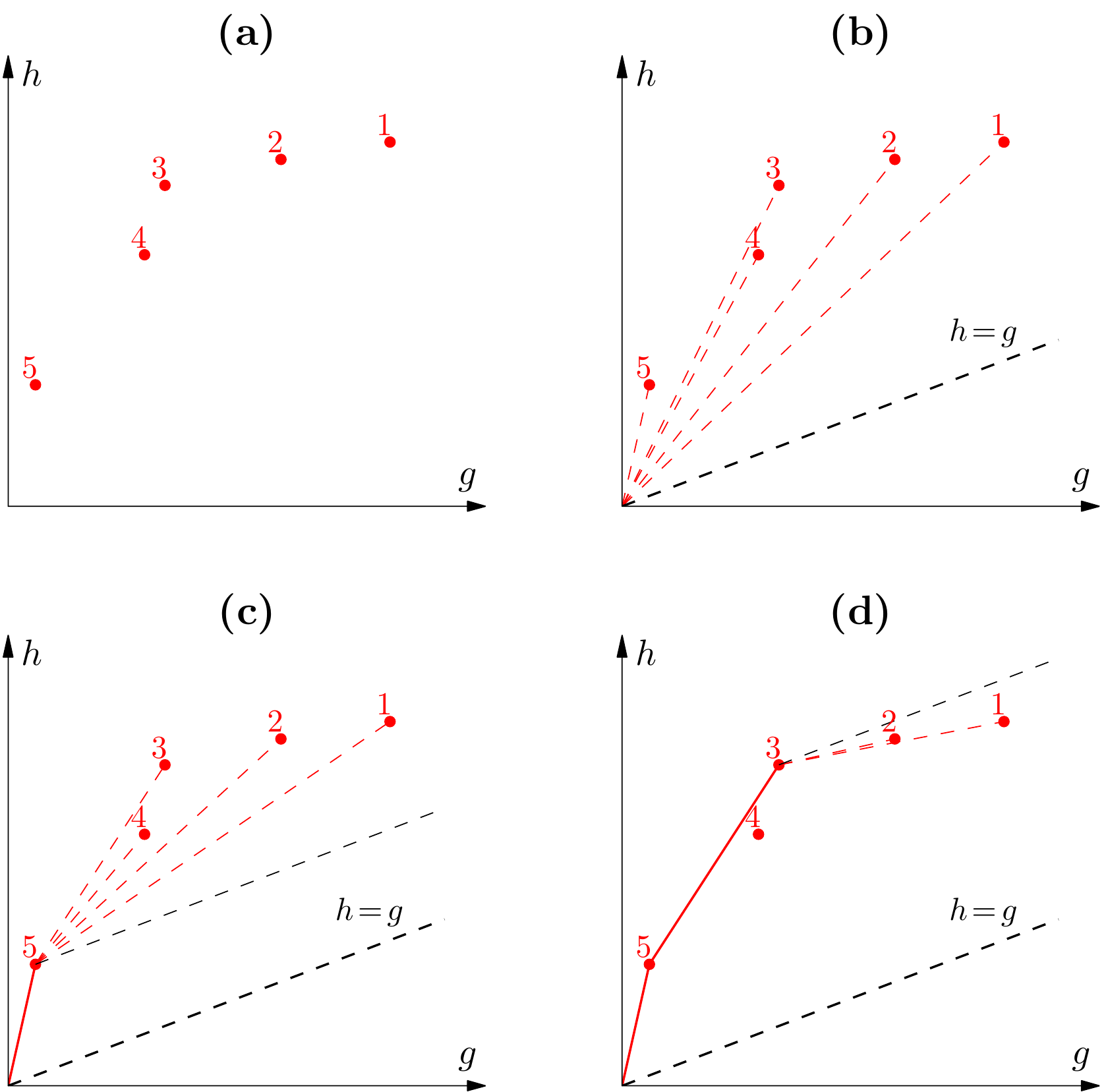}
\end{center}
\caption{Community composition of stable equilibrium can be determined graphically.  Panel~\textbf{a}: In the model each species $i$ has two traits, its stress tolerance $h_i$ and its fecundity $f_i$.  We represent all species in the $h_i$ \textit{vs.} $g_i = \frac{1}{f_i}$ plane (numbered red dots; in the example, there are 5 species).  Panel~\textbf{b}: We start from the origin and draw a line to each of the species (dashed red lines).  The line with the largest slope above the $h=g$ line (dashed black line) corresponds to a species present at equilibrium (in the example, species 5).  Panel~\textbf{c}: To find the other coexisting species, we start from the last found species (in the example, species 5).  We draw a line to each of the species to the right of and above the last found species (dashed red lines).  The line with the largest slope and steeper than the $h=g$ line (dashed black line) corresponds to another species present at equilibrium (in the example, species 3).  Panel~\textbf{d}: We repeat the procedure until there are no lines steeper than the $h=g$ line.  Then we have found all the species coexisting at the stable equilibrium (in the example, species 3 and 5).} \label{fig:intro}
\end{figure}

We have presented a model of an ecological community in which species differ in two traits: stress tolerance and competitive ability.  The analysis of the model equilibria revealed that species coexistence is possible only if the coexisting species satisfy a trade-off between stress tolerance and competitive ability.  We have shown that the coexistence conditions allow a simple graphical representation.  Given a set of species traits, one can easily determine which species are predicted to coexist, as illustrated in Figure~\ref{fig:intro}.  Our model is one of the few mathematical models describing an ecological trade-off with directly verifiable predictions.

Such a verification of the model's predictions in a particular system requires that we know the stress conditions in this system, its regional species pool,  and the stress tolerances and competitive abilities of the species in this species pool. Delineating the species pool and measuring abiotic stress conditions (e.g. pH, temperature, salinity, soil compaction) seem relatively straightforward, but we note that one should not use bioindicators such as Ellenberg indicator values \citep{e79} for this purpose because these are based on the realized communities and these are exactly what we want to predict \citep{zs12}. Determining stress tolerances and competitive abilities may be more difficult to determine. Again, we cannot use realized distributions for this purpose, because we want to establish the species' fundamental rather than the realized niche \citep{h57}. Instead, experiments in monocultures should be conducted, such as \citet{g01} who studied the performance of three marsh species along a salinity gradient, and \citet{j09} who analyzed performance of seven meadow plant species along a flooding gradient.

These studies could be interesting test cases for our model, but also make clear that some assumptions may need to be relaxed for better comparison. Particularly, model~(\ref{eq:discgen}) relies on a strict hierarchy of species where the habitat of less stress tolerant species is contained in the habitat of a more stress tolerant species. Furthermore, model~(\ref{eq:discgen}) assumes that the competitive ability of a species does not depend on the site for which it competes.  The studies cited above show that most species can occur everywhere but their performance strongly depends on abiotic conditions. Our model may still represent a reasonable approximation when one sets a threshold to the abundance of species in order to determine its tolerance.

Alternatively, we can relax the assumptions as follows.  We partition the environment into mutually disjoint habitats $\habi'_k$, $k=1,2,\ldots,H$.  Denoting the size of habitat $\habi'_k$ by $h'_k$, and the competitive ability of species $i$ in habitat $\habi'_k$ by $F_{ik}$, we introduce the model equations
\begin{equation}
 p_i^+ = \sum_{k=1}^{H} h'_k\, L\bigg(\sum_{j=1}^S F_{jk} p_j\bigg)\, F_{ik}p_i.
 \label{eq:compabil}
\end{equation}
The dependence of competitive ability $F_{ik}$ on both species $i$ and habitat $k$ creates the possibility of richer competition scenarios that model~(\ref{eq:discgen}).  Note that model~(\ref{eq:discgen}) can be recovered by setting $H=S$, $\habi'_k = \habi_k\setminus\habi_{k+1}$, $h'_k = h_k-h_{k+1}$ and
\[
 F_{ik} = \begin{cases} f_i & \text{if $i \geq k$} \\
 0 & \text{if $i<k$.} \end{cases}
\]
\citet{d13} have explored the consequences of a model that retains the strict hierarchy but allows competitive ability to be stress-dependent; this model excludes systems where species have adapted to certain abiotic conditions and their performance is bell-shaped. \citet{d13} focus primarily on the the question whether theoretically there can be an infinite number of coexisting species, and show that this is no longer possible under their model. Although these results are theoretically intriguing, they are less relevant for predicting which species from a finite species pool may coexist and at what abundances. A (graphical) tool such as we present in this paper is very desirable, but probably unfeasible for this more complex model. One possible avenue of future research is to study under which conditions our graphical analysis provides a reasonable approximation to this problem.

When formulating model~(\ref{eq:discgen}), we have simplified the spatial structure of the model as much as possible, as illustrated in Figure~\ref{fig:spatial}.  It is relatively easy to reintroduce some of the spatial complexity into the model equations.  We consider non-uniform, species-dependent dispersal.  As for model~(\ref{eq:compabil}), we consider a partition of the environment into $H$ habitats $\habi'_k$ with size $h'_k$.  As for model~(\ref{eq:contfull}), we augment the set of dynamical variables and consider the occupancy $p_{ik}$  of species $i$ in habitat $\habi'_k$.  The competitive ability $F_{i\ell k}$ depend on species $i$, the habitat $\habi'_\ell$ from which it disperses and the habitat $\habi'_k$ in which it tries to establish.  The model equations are
\begin{equation}
 p_{ik}^+ = h'_k\,
 L\bigg(\sum_{j=1}^S \sum_{\ell=1}^H F_{j\ell k} p_{j\ell}\bigg)
 \sum_{\ell=1}^H F_{i\ell k} p_{i\ell}.
\end{equation}
Model~(\ref{eq:compabil}) can be recovered by taking competitive abilities $F_{i\ell k}$ independent of index $\ell$, that is, independent of the habitat a species is dispersing from.

Other extensions of the model include a fully spatially explicit approach and consideration of trade-offs between more than two species. For all such ventures, a simple null model is convenient to assess the relevance of these extensions. We hope to have provided such a simple null model where our graphical tool greatly facilitates such analyses and makes these models accessible to a less mathematical audience.

\section*{Acknowledgements}

We thank S.~Schreiber and two anonymous reviewers for constructive comments.  BH thanks L.~Gironne for a helpful suggestion.  Financial support was provided for BH by the TULIP Laboratory of Excellence (ANR-10-LABX-41), for RSE by a VIDI grant from the Netherlands Scientific Organization (NWO) and for BH and RSE by an exchange grant from the Van Gogh programme.

\newpage
\section*{Appendix A: Recruitment probability} 

Here we derive the probability that species $i$ in competition with $n$ other species recruits a vacant site.  The derivation, which is based on standard properties of Poisson random variables, is not new, but seems to be unfamiliar to many ecologists.

We consider a vacant site.  The number of propagules (e.g., seeds) of species $i$ reaching the vacant site is modelled by a Poisson random variable.  We denote this number by $s_i$.  The parameter of the Poisson random variable is equal to $f_i p_i$.  That is, the probability of dispersing $s_i$ seeds to the vacant site is
\[
 P(s_i;f_i p_i) = \e^{-f_ip_i} \frac{(f_ip_i)^{s_i}}{s_i!}.
\]
Species $i$ competes with $n$ other species.  We label these species by $j_1,j_2,\ldots,j_n$.  Conditional on the number of propagules $s_i, s_{j_1}, \ldots, s_{j_n}$, the probability that species $i$ wins the competition for the vacant site is
\begin{align*}
 &\Prob\big(\text{species $i$ occupies vacant site}\,\big|\,s_i,(s_{j_k})\big) \\
 &\qquad\qquad = \begin{cases}
 \frac{\ts s_i}{\ts s_i + \sum_k s_{j_k}} & \text{if $s_i > 0$} \\[8pt]
 0 & \text{if $s_i = 0$}.
 \end{cases}
\end{align*}
We average this expression with respect to the Poisson distributions for $s_i$ and $s_{j_k}$,
\begin{align*}
 &\Prob\big(\text{species $i$ occupies vacant site}\big) \\
 &\qquad = \sum_{s_i=1}^\infty P(s_i;f_ip_i)
 \sum_{(s_{j_k})} \left( \prod_k P(s_{j_k};f_{j_k}p_{j_k}) \right)
 \frac{s_i}{s_i + \sum_k s_{j_k}} \\
 &\qquad = \sum_{s_i=1}^\infty P(s_i;f_ip_i)
 \sum_{s_j=0}^\infty P(s_j;{\ts \sum_{k} f_{j_k}p_{j_k}})
 \frac{s_i}{s_i + s_j} \\
 &\qquad = f_i p_i \sum_{s_i=1}^\infty \e^{-f_ip_i}
 \frac{(f_ip_i)^{s_i-1}}{(s_i-1)!}
 \sum_{s_j=0}^\infty P(s_j;{\ts \sum_{k} f_{j_k}p_{j_k}})
 \frac{1}{s_i + s_j} \\
 &\qquad = f_i p_i \sum_{s_i=0}^\infty P(s_i;f_ip_i)
 \sum_{s_j=0}^\infty P(s_j;{\ts \sum_{k} f_{j_k}p_{j_k}})
 \frac{1}{s_i + s_j + 1} \\
 &\qquad = f_i p_i \sum_{s=0}^\infty P(s;f_ip_i +
 {\ts \sum_{k} f_{j_k}p_{j_k}}) \frac{1}{s + 1}
\end{align*}
where we have used that the sum of Poisson random variables is again a Poisson random variable in the second equality and in the last equality.  Because
\begin{align*}
 \sum_{s=0}^\infty P(s;\lambda) \frac{1}{s + 1}
 &= \e^{-\lambda} \sum_{s=0}^\infty \frac{\lambda^s}{(s+1)!} \\
 &= \frac{\e^{-\lambda}}{\lambda} \sum_{s=0}^\infty \frac{\lambda^{s+1}}{(s+1)!} \\
 &= \frac{\e^{-\lambda}}{\lambda} \big(\e^\lambda - 1\big) \\
 &= \frac{1 - \e^{-\lambda}}{\lambda} \\
 &= L(\lambda),
\end{align*}
where the last line is the definition of the function $L$, we have
\begin{align*}
 &\Prob\big(\text{species $i$ occupies vacant site}\big) \\
 &\qquad = f_i p_i \, L\big({\ts f_ip_i + \sum_{k} f_{j_k}p_{j_k}}\big).
\end{align*}

The probability that no species occupies the vacant site, that is, the probability that the vacant site remains empty, is
\begin{align*}
 &\Prob\big(\text{no species occupies vacant site}\big) \\[6pt]
 &= P(0;f_ip_i) \prod_k P(0;f_{j_k}p_{j_k}) \\[-3pt]
 &= e^{-\big(f_ip_i + \sum_{k} f_{j_k}p_{j_k}\big)}.
\end{align*}

\section*{Appendix B: Proof of Result 1} 

We consider a subset $\sC$ of the set of all species $\sS$, and study whether there exists an equilibrium of dynamical system~(\ref{eq:discgen}) in which the species in $\sC$ are present and the species in $\sS\setminus\sC$ are absent.  We write the dynamical system as
\begin{subequations} \label{eq:discgenrewr}
\begin{equation}
 p_i^+ = F_i(\tpl{p}) f_i p_i
 \qquad \text{$i\in\sS$},
\end{equation}
with
\begin{equation} \label{eq:defF}
 F_i(\tpl{p}) = \sum_{n=i}^{S-1} (h_n-h_{n+1})\,
 L\Big( \sum_{m=1}^n f_m p_m \Big)
 + h_S\, L\Big( \sum_{m=1}^S f_m p_m \Big)
\end{equation}
\end{subequations}
and
\begin{equation*}
 L(x) = \frac{1-\e^{-x}}{x}.
\end{equation*}

An equilibrium $\tpl{\widetilde p}$ of (\ref{eq:discgenrewr}) has to satisfy
\begin{equation}
 \widetilde p_i = F_i(\tpl{\widetilde p}) f_i \widetilde p_i
 \qquad \text{for all $i\in\sS$}.
 \label{eq:equicond}
\end{equation}
Equation~(\ref{eq:equicond}) is trivially satisfied for $i \in \sS\setminus\sC$, that is, for species that are absent at equilibrium ($\widetilde p_i = 0$).  For species that are present at equilibrium ($\widetilde p_i > 0$), equation~(\ref{eq:equicond}) simplifies to
\begin{equation}
 F_{i_k}(\tpl{\widetilde p}) = g_{i_k}
 \qquad \text{for all $k=1,2,\ldots,C$}.
 \label{eq:equicondred}
\end{equation}

\subsubsection*{Preliminary computation}

We derive an expression for $F_i(\tpl{\widetilde p})$.  As we will need a more general result below, we consider the quantity $G_i$ which is equal to $F_i(\tpl{\widetilde p})$
in which the function $L$ is replaced by an arbitrary function $K$,
\begin{equation}
 G_i = \sum_{n=i}^{S-1} (h_n-h_{n+1})\,
 K\Big( \sum_{m=1}^n f_m \widetilde p_m \Big)
 + h_S\, K\Big( \sum_{m=1}^S f_m \widetilde p_m \Big).
\end{equation}

For a species $i$ with $i < i_C$, we determine the index $k$ such that $i_{k} \leq i < i_{k+1}$ with $k \in \{1,2,\ldots,C-1\}$ and set $k=0$ if $i < i_1$.  We prove that
\begin{equation}
 G_i = \begin{cases}
 \begin{aligned}
 & (h_i-h_{i_{k+1}})\, K\Big( \sum_{j=1}^{k} f_{i_j}
   \widetilde p_{i_j} \Big) \\[-6pt]
 & + \sum_{\ell=k+1}^{C-1} (h_{i_\ell}-h_{i_{\ell+1}})\,
   K\Big( \sum_{j=1}^\ell f_{i_j} \widetilde p_{i_j} \Big) \\[-4pt]
 & + h_{i_C}\, K\Big( \sum_{j=1}^C f_{i_j} \widetilde p_{i_j} \Big)
 \end{aligned}
 & \qquad\text{if $i < i_C$} \\[50pt]
 \ds h_i\, K\Big( \sum_{j=1}^C f_{i_j} \widetilde p_{i_j} \Big)
 & \qquad\text{if $i \geq i_C$}.
 \end{cases}
 \label{eq:equicondreda}
\end{equation}

For $i < i_C$ we determine the index $k$ as above and we compute
\begin{align*}
 F_i(\tpl{\widetilde p})
 &= \sum_{n=i}^{S-1} (h_n-h_{n+1})\,
    K\Big( \sum_{m=1}^n f_m \widetilde p_m \Big)
  + h_S\, K\Big( \sum_{m=1}^S f_m \widetilde p_m \Big) \\
 &= \sum_{n=i}^{i_{k+1}-1} (h_n-h_{n+1})\,
    K\Big( \sum_{m=1}^n f_m \widetilde p_m \Big) \\ &\quad+ 
    \sum_{\ell=k+1}^{C-1} \sum_{n=i_\ell}^{i_{\ell+1}-1} (h_n-h_{n+1})\,
    K\Big( \sum_{m=1}^n f_m \widetilde p_m \Big) \\ &\quad+
    \sum_{n=i_C}^{S-1} (h_n-h_{n+1})\,
    K\Big( \sum_{m=1}^n f_m \widetilde p_m \Big)
  + h_S\, K\Big( \sum_{m=1}^S f_m \widetilde p_m \Big) \\
 &= \sum_{n=i}^{i_{k+1}-1} (h_n-h_{n+1})\,
    K\Big( \sum_{j=1}^{k} f_{i_j} \widetilde p_{i_j} \Big) \\ &\quad+ 
    \sum_{\ell=k+1}^{C-1} \sum_{n=i_\ell}^{i_{\ell+1}-1} (h_n-h_{n+1})\,
    K\Big( \sum_{j=1}^\ell f_{i_j} \widetilde p_{i_j} \Big) \\ &\quad+
    \sum_{n=i_C}^{S-1} (h_n-h_{n+1})\,
    K\Big( \sum_{j=1}^C f_{i_j} \widetilde p_{i_j} \Big)
  + h_S\, K\Big( \sum_{j=1}^C f_{i_j} \widetilde p_{i_j} \Big) \\
 &= \bigg( \sum_{n=i}^{i_{k+1}-1} (h_n-h_{n+1}) \bigg)\,
    K\Big( \sum_{j=1}^{k} f_{i_j} \widetilde p_{i_j} \Big) \\ &\quad+ 
    \sum_{\ell=k+1}^{C-1} \bigg( \sum_{n=i_\ell}^{i_{\ell+1}-1} (h_n-h_{n+1}) \bigg)\,
    K\Big( \sum_{j=1}^\ell f_{i_j} \widetilde p_{i_j} \Big) \\ &\quad+
    \bigg( \sum_{n=i_C}^{S-1} (h_n-h_{n+1}) + h_S \bigg) \,
    K\Big( \sum_{j=1}^C f_{i_j} \widetilde p_{i_j} \Big) \\
 &= (h_i-h_{i_{k+1}})\,
    K\Big( \sum_{j=1}^{k} f_{i_j} \widetilde p_{i_j} \Big) \\ &\quad+
    \sum_{\ell=k+1}^{C-1} (h_{i_\ell}-h_{i_{\ell+1}})\,
    K\Big( \sum_{j=1}^\ell f_{i_j} \widetilde p_{i_j} \Big) \\ &\quad+
    h_{i_C}\, K\Big( \sum_{j=1}^C f_{i_j} \widetilde p_{i_j} \Big).
\end{align*}
A similar computation holds for $i \geq i_C$, proving (\ref{eq:equicondreda}).

\subsubsection*{Equilibrium conditions}

Putting $K=L$ and $i=i_k$ in (\ref{eq:equicondreda}) we get
\begin{equation}
 F_{i_k}(\tpl{\widetilde p}) = 
 \sum_{\ell=k}^{C-1} (h_{i_\ell}-h_{i_{\ell+1}})\,
 L\Big( \sum_{j=1}^\ell f_{i_j} \widetilde p_{i_j} \Big) +
 h_{i_C}\, L\Big( \sum_{j=1}^C f_{i_j} \widetilde p_{i_j} \Big).
\end{equation}
Hence, the equilibrium conditions~(\ref{eq:equicondred}) read
\begin{subequations} \label{eq:equicondredb}
\begin{align}
 \sum_{\ell=k}^{C-1} (h_{i_\ell}-h_{i_{\ell+1}})\,
   L\Big( \sum_{j=1}^\ell f_{i_j} \widetilde p_{i_j} \Big)
 + h_{i_C}\, L\Big( \sum_{j=1}^C f_{i_j} \widetilde p_{i_j} \Big)
 &= g_{i_k} \notag \\[-6pt]
 & \hspace{-4ex} \text{for all $k=1,2,\ldots,C-1$} \\
 h_{i_C}\, L\Big( \sum_{j=1}^C f_{i_j} \widetilde p_{i_j} \Big)
 &= g_{i_C}.
\end{align}
\end{subequations}
Note that the parameters $h_i$ and $f_i$ with $i \in \sS\setminus\sC$ have been eliminated from (\ref{eq:equicondredb}).

Rearranging (\ref{eq:equicondredb}) we get
\begin{subequations} \label{eq:equicondredc}
\begin{align}
 & L\Big( \sum_{j=1}^k f_{i_j} \widetilde p_{i_j} \Big)
 = \frac{g_{i_k}-g_{i_{k+1}}}{h_{i_k}-h_{i_{k+1}}}
 \qquad \text{for all $k=1,2,\ldots,C-1$} \\
 & L\Big( \sum_{j=1}^C f_{i_j} \widetilde p_{i_j} \Big)
 = \frac{g_{i_C}}{h_{i_C}}.
\end{align}
\end{subequations}
We have to invert the function $L$ to solve for the occupancies $\widetilde p_i$.  Because $L$ maps the positive real line $(0,\infty)$ to the real interval $(0,1)$, the inversion is possible if and only if
\begin{subequations} \label{eq:condexist1}
\begin{align}
 0 < \frac{g_{i_k}-g_{i_{k+1}}}{h_{i_k}-h_{i_{k+1}}} &< 1
 \qquad \text{for all $k=1,2,\ldots,C-1$} \\
 \frac{g_{i_C}}{h_{i_C}} &< 1,
\end{align}
\end{subequations}
where we have used Assumption~\ref{ass:generic} to discard parameter combinations for which one or more of these inequalities is replaced by an equality.

After inverting the function $L$, the solutions for the occupancies $\widetilde p_i$ should be positive.  Because $L$ is decreasing, this condition is satisfied if and only if
\begin{equation}
 \frac{g_{i_C}}{h_{i_C}} 
 < \frac{g_{i_{C-1}}-g_{i_C}}{h_{i_{C-1}}-h_{i_C}}
 < \ldots
 < \frac{g_{i_2}-g_{i_3}}{h_{i_2}-h_{i_3}}
 < \frac{g_{i_1}-g_{i_2}}{h_{i_1}-h_{i_2}},
 \label{eq:condexist2}
\end{equation}
where we have used Assumption~\ref{ass:generic} to discard cases in which equality rather than inequality holds.

Combining (\ref{eq:condexist1}) and (\ref{eq:condexist2}) we get the following conditions:
\begin{equation}
 \frac{g_{i_C}}{h_{i_C}} 
 < \frac{g_{i_{C-1}}-g_{i_C}}{h_{i_{C-1}}-h_{i_C}}
 < \ldots
 < \frac{g_{i_2}-g_{i_3}}{h_{i_2}-h_{i_3}}
 < \frac{g_{i_1}-g_{i_2}}{h_{i_1}-h_{i_2}} < 1,
 \label{eq:condexist}
\end{equation}
which proves the first statement of Result~\ref{res:equilib}.

If these conditions are satisfied, the equilibrium occupancies $\widetilde p_i$ are given by
\begin{subequations} \label{eq:equioccu}
\begin{align}
 & \sum_{j=1}^k f_{i_j} \widetilde p_{i_j}
 = L^{-1}\Big( \frac{g_{i_k}-g_{i_{k+1}}}{h_{i_k}-h_{i_{k+1}}} \Big)
 \qquad \text{$k=1,2,\ldots,C-1$} \\
 & \sum_{j=1}^C f_{i_j} \widetilde p_{i_j}
 = L^{-1}\Big( \frac{g_{i_C}}{h_{i_C}} \Big),
\end{align}
\end{subequations}
with $L^{-1}$ the inverse function of $L$.  Rewriting equations~(\ref{eq:equioccu}) we get
\begin{subequations} \label{eq:laststep}
\begin{align}
 \widetilde p_{i_1} &=
 g_{i_1} \; L^{-1}\bigg(\frac{g_{i_{1}}-g_{i_{2}}}{h_{i_{1}}-h_{i_{2}}}\bigg) \\
 \widetilde p_{i_k} &=
 g_{i_k} \Bigg( L^{-1}\bigg(\frac{g_{i_{k}}-g_{i_{k+1}}}{h_{i_{k}}-h_{i_{k+1}}}\bigg)
 - L^{-1}\bigg(\frac{g_{i_{k-1}}-g_{i_{k}}}{h_{i_{k-1}}-h_{i_{k}}}\bigg) \Bigg)
 \qquad \text{$k = 2,3,\ldots,C-1$} \\
 \widetilde p_{i_C} &=
 g_{i_C} \Bigg( L^{-1}\bigg(\frac{g_{i_{C}}}{h_{i_{C}}}\bigg)
 - L^{-1}\bigg(\frac{g_{i_{C-1}}-g_{i_{C}}}{h_{i_{C-1}}-h_{i_{C}}}\bigg) \Bigg).
\end{align}
\end{subequations}

Recall the definition $L(x) = \frac{1-\e^{-x}}{x}$.  The function $\Lambda$ is defined as the inverse function of $x \mapsto \frac{x}{1-\e^{-x}}$.  We have \\[-3pt]
\begin{equation*}
 y=\Lambda(x)
  \ \Leftrightarrow\ x = \frac{1}{L(y)} 
  \ \Leftrightarrow\ \frac{1}{x} = L(y)
  \ \Leftrightarrow\ y = L^{-1}\big(\frac{1}{x}\big).
\end{equation*}
Hence, $L^{-1}(x) = \Lambda\big(\frac{1}{x}\big)$.  Using this identity in (\ref{eq:laststep}), we obtain the second statement of Result~\ref{res:equilib}.

The function $\Lambda$ can be expressed in terms of the Lambert $W$ function.  We have
\begin{align*}
 y=\Lambda(x)
  &\ \Leftrightarrow\  x = \frac{y}{1-\e^{-y}} = \frac{y\,\e^y}{\e^y-1} \\
  &\ \Leftrightarrow\  y\,\e^y = x\,\big(\e^y-1\big) \\
  &\ \Leftrightarrow\ (y-x)\,\e^y= -x \\
  &\ \Leftrightarrow\ (y-x)\,\e^{y-x} = -x\,\e^{-x} \\
  &\ \Leftrightarrow\ y-x = W_0(-x\,\e^{-x}),
\end{align*}
with $W_0$ the upper branch of the Lambert $W$ function.  Hence,
\begin{equation}
 \Lambda(x) = x + W_0(-x\,\e^{-x}).
 \label{eq:lamb1}
\end{equation}
Similarly, for the function $L^{-1}$,
\begin{align*}
 y=L^{-1}(x)
  &\ \Leftrightarrow\  x = \frac{1-\e^{-y}}{y} = \frac{\e^y-1}{y\,\e^y} \\
  &\ \Leftrightarrow\  y\,\e^y = \frac{1}{x}\,\big(\e^y-1\big) \\
  &\ \Leftrightarrow\  \Big(y-\frac{1}{x}\Big)\,\e^y = -\frac{1}{x} \\
  &\ \Leftrightarrow\  \Big(y-\frac{1}{x}\Big)\,\e^{y-\frac{1}{x}} = -\frac{1}{x}\,\e^{-\frac{1}{x}} \\
  &\ \Leftrightarrow\  y-\frac{1}{x} = W_0\Big(-\frac{1}{x}\,\e^{-\frac{1}{x}}\Big),
\end{align*} 
so that
\begin{equation}
 L^{-1}(x) = \frac{1}{x} + W_0\Big(-\frac{1}{x}\,\e^{-\frac{1}{x}}\Big).
 \label{eq:lamb2}
\end{equation}
Note that equations (\ref{eq:lamb1}) and (\ref{eq:lamb2}) satisfy $L^{-1}(x) = \Lambda\big(\frac{1}{x}\big)$.

Because we have derived an explicit expression for the equilibrium $\tpl{\widetilde p}$, we have also established the third statement of Result~\ref{res:equilib}.

\section*{Appendix C: Proof of Result 2} 

We consider an equilibrium $\tpl{\widetilde p}$ of dynamical system~(\ref{eq:discgen}) and study its local stability.  We denote by $\sC$ the subset of the set of all species $\sS$ that are present in the equilibrium $\tpl{\widetilde p}$. 

The equilibrium $\tpl{\widetilde p}$ is locally asymptotically stable if and only if the eigenvalues of the Jacobian matrix evaluated at $\tpl{\widetilde p}$ lie inside the unit circle of the complex plane.  Using the notation of (\ref{eq:discgenrewr}), the Jacobian matrix $J$ evaluated at $\tpl{\widetilde p}$ is
\begin{align}
 J = \quad &\begin{pmatrix}
 f_1 F_1(\tpl{\widetilde p}) & 0 & \cdots & 0 \\[4pt]
 0 & f_2 F_2(\tpl{\widetilde p}) & \cdots & 0 \\[-2pt]
 \vdots & \vdots & & \vdots \\[2pt]
 0 & 0 & \cdots & f_S F_S(\tpl{\widetilde p})
 \end{pmatrix} \notag \\[6pt] + &\begin{pmatrix}
 f_1\widetilde p_1\,\dd{F_1}{p_1}(\tpl{\widetilde p}) &
 f_1\widetilde p_1\,\dd{F_1}{p_2}(\tpl{\widetilde p}) & \cdots &
 f_1\widetilde p_1\,\dd{F_1}{p_S}(\tpl{\widetilde p}) \\[4pt]
 f_2\widetilde p_2\,\dd{F_2}{p_1}(\tpl{\widetilde p}) &
 f_2\widetilde p_2\,\dd{F_2}{p_2}(\tpl{\widetilde p}) & \cdots &
 f_2\widetilde p_2\,\dd{F_2}{p_S}(\tpl{\widetilde p}) \\
 \vdots & \vdots & & \vdots \\[4pt]
 f_S\widetilde p_S\,\dd{F_S}{p_1}(\tpl{\widetilde p}) &
 f_S\widetilde p_S\,\dd{F_S}{p_2}(\tpl{\widetilde p}) & \cdots &
 f_S\widetilde p_S\,\dd{F_S}{p_S}(\tpl{\widetilde p}) 
 \end{pmatrix}.
 \label{eq:jacobian1}
\end{align}

Consider a species $j$ that is absent at equilibrium, $j \in \sS\setminus\sC$ and $\widetilde p_j = 0$.  Row $j$ of the second term of (\ref{eq:jacobian1}) is zero, so that row $j$ of the Jacobian matrix $J$ is zero except for the diagonal element $f_j F_j(\tpl{\widetilde p})$.  Hence, by applying the permutation $P$ that shifts the indices in $\sC$ in front of the indices in $\sS\setminus\sC$, we get a block matrix
\begin{equation}
 P^\top J P = \begin{pmatrix} J_\sC & K \\[4pt] 0 & J_{\sS\setminus\sC} \end{pmatrix},
 \label{eq:jacobian2}
\end{equation}
with $J_\sC$ a $C \times C$ matrix, $K$ a $C \times (S\!-\!C)$ matrix, and $J_{\sS\setminus\sC}$ a diagonal $(S\!-\!C) \times (S\!-\!C)$ matrix.  It follows that the eigenvalues of $J$ are equal to the union of the eigenvalues of $J_\sC$ and the eigenvalues of $J_{\sS\setminus\sC}$.

\subsubsection*{Eigenvalues of $J_{\sS\setminus\sC}$}

Because the matrix $J_{\sS\setminus\sC}$ is diagonal, its eigenvalues are given by its diagonal elements.  The diagonal elements are equal to $f_j F_j(\tpl{\widetilde p})$ with $j \in \sS\setminus\sC$, implying that the eigenvalues of $J_{\sS\setminus\sC}$ are real and positive.  To obtain the stability conditions, we look for conditions equivalent with $f_j F_j(\tpl{\widetilde p}) < 1$ for all $j \in \sS\setminus\sC$, that is, for all $j \in \sS$ excluding $i_1,i_2,\ldots,i_C$.

First, we consider a species $j$ for which $i_k < j < i_{k+1}$ with $k = 1,2,\ldots C-1$.  Expression~(\ref{eq:equicondreda}) with $K=L$ and $i=j$ gives
\begin{align*}
 F_j(\tpl{\widetilde p}) &= (h_j-h_{i_{k+1}})\,
 L\Big( \sum_{m=1}^k f_{i_m} \widetilde p_{i_m} \Big) \\[-4pt] & \quad+
 \sum_{\ell=k+1}^{C-1} (h_{i_\ell}-h_{i_{\ell+1}})\,
 L\Big( \sum_{m=1}^\ell f_{i_m} \widetilde p_{i_m} \Big) +
 h_{i_C}\, L\Big( \sum_{m=1}^C f_{i_m} \widetilde p_{i_m} \Big).
\end{align*}
We use equation~(\ref{eq:equicondredc}a) for the first term of the right-hand side and equation~(\ref{eq:equicondredb}a) for the second and third term of the right-hand side.  This leads to
\[
 F_j(\tpl{\widetilde p}) = (h_j-h_{i_{k+1}})\,
 \frac{g_{i_k}-g_{i_{k+1}}}{h_{i_k}-h_{i_{k+1}}} + g_{i_{k+1}},
\]
so that the condition $f_j F_j(\tpl{\widetilde p}) < 1$ is equivalent with
\[
 (h_j-h_{i_{k+1}})\,
 \frac{g_{i_k}-g_{i_{k+1}}}{h_{i_k}-h_{i_{k+1}}}
 < g_j - g_{i_{k+1}}.
\]
Using that $h_{i_k} > h_j > h_{i_{k+1}}$ (because $i_k < j < i_{k+1}$), we see that the latter condition is equivalent with
\begin{subequations} \label{eq:eigJSC}
\begin{equation}
 \frac{g_{i_k}-g_j}{h_{i_k}-h_j} <
 \frac{g_j-g_{i_{k+1}}}{h_j-h_{i_{k+1}}}.
\end{equation}

Second, we consider a species $j$ for which $j < i_1$.  Expression~(\ref{eq:equicondreda}) with $K=L$, $i=j$ and $k=0$ gives
\[
 F_j(\tpl{\widetilde p}) = (h_j-h_{i_1})\,L(0) +
 \sum_{\ell=1}^{C-1} (h_{i_\ell}-h_{i_{\ell+1}})\,
 L\Big( \sum_{m=1}^\ell f_{i_m} \widetilde p_{i_m} \Big) +
 h_{i_C}\, L\Big( \sum_{m=1}^C f_{i_m} \widetilde p_{i_m} \Big).
\]
We use $L(0)=1$ for the first term and equation~(\ref{eq:equicondredb}a) for the second and third term of the right-hand side.  This leads to
\[
 F_j(\tpl{\widetilde p}) = h_j-h_{i_1} + g_{i_1},
\]
so that the condition $f_j F_j(\tpl{\widetilde p}) < 1$ is equivalent with
\[
 h_j-h_{i_1} < g_j - g_{i_1}.
\]
Using that $h_j > h_{i_1}$ (because $j < i_1$), we see that the latter condition is equivalent with
\begin{equation}
 1 < \frac{g_j-g_{i_1}}{h_j-h_{i_1}}.
\end{equation}

Third, we consider a species $j$ for which $i_C < j$.  Expression (\ref{eq:equicondreda}) with $K=L$, $i=j$ and $k=C$ gives
\[
 F_j(\tpl{\widetilde p})
 = h_{j}\, L\Big( \sum_{m=1}^C f_{i_m} \widetilde p_{i_m} \Big)
 = h_{j}\, \frac{g_{i_C}}{h_{i_C}},
\]
where we have used equation (\ref{eq:equicondredb}b).  Hence, the condition $f_j F_j(\tpl{\widetilde p}) < 1$ is equivalent with
\[
 h_j \frac{g_{i_C}}{h_{i_C}} < g_j.
\]
Using that $h_{i_C} > h_j$ (because $i_C < j$), we see that the latter condition is equivalent with
\begin{equation}
 \frac{g_{i_C}-g_j}{h_{i_C}-h_j} < \frac{g_j}{h_j}.
\end{equation}
\end{subequations}

The set of conditions (\ref{eq:eigJSC}) for all $j \in \sS\setminus\sC$ is equivalent with conditions (\ref{eq:condstable}).  Note that Assumption~\ref{ass:generic} allows us to discard parameter combinations for which one or more of these inequalities is replaced by an equality.  Hence, when one of the conditions (\ref{eq:condstable}) is violated, the matrix $J_{\sS\setminus\sC}$ has an eigenvalue $\lambda>1$ and the equilibrium $\tpl{\widetilde p}$ is unstable.  As a result, we have proved that conditions (\ref{eq:condstable}) are necessary for local stability.  To prove that conditions (\ref{eq:condstable}) are also sufficient for local stability, it suffices to prove that the eigenvalues of $J_\sC$ lie inside the unit circle.

\subsubsection*{Eigenvalues of $J_\sC$}

The matrix $J_\sC$ is
\begin{equation*}
 J_\sC = \idty + \begin{pmatrix}
 f_{i_1}\widetilde p_{i_1}\,\dd{F_{i_1}}{p_{i_1}}(\tpl{\widetilde p}) &
 f_{i_1}\widetilde p_{i_1}\,\dd{F_{i_1}}{p_{i_2}}(\tpl{\widetilde p}) &
 \cdots &
 f_{i_1}\widetilde p_{i_1}\,\dd{F_{i_1}}{p_{i_C}}(\tpl{\widetilde p}) \\[4pt]
 f_{i_2}\widetilde p_{i_2}\,\dd{F_{i_2}}{p_{i_1}}(\tpl{\widetilde p}) &
 f_{i_2}\widetilde p_{i_2}\,\dd{F_{i_2}}{p_{i_2}}(\tpl{\widetilde p}) &
 \cdots &
 f_{i_2}\widetilde p_{i_2}\,\dd{F_{i_2}}{p_{i_C}}(\tpl{\widetilde p}) \\
 \vdots & \vdots & & \vdots \\[4pt]
 f_{i_C}\widetilde p_{i_C}\,\dd{F_{i_C}}{p_{i_1}}(\tpl{\widetilde p}) &
 f_{i_C}\widetilde p_{i_C}\,\dd{F_{i_C}}{p_{i_2}}(\tpl{\widetilde p}) &
 \cdots &
 f_{i_C}\widetilde p_{i_C}\,\dd{F_{i_C}}{p_{i_C}}(\tpl{\widetilde p})
 \end{pmatrix},
\end{equation*}
with $\idty$ the identity matrix.

We compute the partial derivatives,
\[
 \dd{F_i}{p_j}(\tpl{p}) = 
 \sum_{n=\max(i,j)}^{S-1} (h_n-h_{n+1})\,
 L'\Big( \sum_{m=1}^n f_m p_m \Big) f_j
 + h_S\, L'\Big( \sum_{m=1}^S f_m p_m \Big) f_j.
\]
Using (\ref{eq:equicondreda}) with $K=L'$ and $i = \max(i_k,i_m) = i_{\max(k,m)}$, we get
\[
 \dd{F_{i_k}}{p_{i_m}}(\tpl{\widetilde p})
 = \sum_{\ell=\max(k,m)}^{C-1} (h_{i_\ell}-h_{i_{\ell+1}})\,
 L'\Big( \sum_{j=1}^\ell f_{i_j} \widetilde p_{i_j} \Big) f_{i_m}
 + h_{i_C}\, L'\Big( \sum_{j=1}^C f_{i_j} \widetilde p_{i_j} \Big) f_{i_m}.
\]
Note that the parameters $h_i$ and $f_i$ with $i \in \sS\setminus\sC$ do not appear in these partial derivatives.

We can write the matrix $J_\sC$ as
\begin{equation*}
 J_\sC = \idty - G H,
\end{equation*}
with $G$ a diagonal matrix,
\begin{equation}
 G = \begin{pmatrix}
 \widetilde p_{i_1} & 0 & \cdots & 0 \\[4pt]
 0 & \widetilde p_{i_2} & \cdots & 0 \\[-2pt]
 \vdots & \vdots & & \vdots \\[2pt]
 0 & 0 & \cdots & \widetilde p_{i_C}
 \end{pmatrix},
 \label{eq:matrixG}
\end{equation}
and $H$ a symmetric matrix,
\begin{equation}
 H = \sum_{k=1}^{C-1} (h_{i_k}-h_{i_{k+1}})
 \; \Big| L'\Big( \sum_{j=1}^k f_{i_j}\widetilde p_{i_j} \Big) \Big|
 \; \tpl{v}_k \tpl{v}_k^\top + \; h_{i_C}
 \, \Big| L'\Big( \sum_{j=1}^C f_{i_j}\widetilde p_{i_j} \Big) \Big|
 \; \tpl{v}_C \tpl{v}_C^\top,
 \label{eq:matrixH}
\end{equation}
where we have used that $L'(x) = -|L'(x)|$ because $L$ is decreasing.  The vectors $\tpl{v}_k$ are
\begin{equation}
 \tpl v_k = \sum_{j=1}^k f_{i_j}\;\tpl e_j,
 \label{eq:setvect1}
\end{equation}
where the vectors $\tpl e_j$ are the standard basis vectors in $\R^C$.

First we establish that all eigenvalues of the matrix $G H$ are real and positive.  Then we show that all eigenvalues of the matrix $G H$ lie inside the unit circle.  As a result, all eigenvalues of $J_\sC = \idty - GH$ belong to the real interval $[0,1)$.

To show that all eigenvalues of the matrix $G H$ are real and positive, we first consider the matrix $H$,
\[
 H = \sum_{k=1}^C a_k\; \tpl{v}_k \tpl{v}_k^\top,
\]
a linear combination of symmetric rank-one matrices $\tpl{v}_k \tpl{v}_k^\top$, with vectors $\tpl{v}_k$ given by (\ref{eq:setvect1}) and coefficients $a_k$ given by (\ref{eq:matrixH}). Because the coefficients $a_k$ are positive, the matrix $H$ is a sum of positive-semidefinite matrices, and therefore positive-semidefinite.

In fact, the matrix $H$ is positive-definite.  To establish this, we have to prove that $H$ is not singular.  Assume that $H$ is singular.  Then there exists a vector $\tpl{x} \neq \tpl{0}$ for which $H\tpl{x} = \tpl{0}$.  Then also $\tpl{x}^\top H\tpl{x} = 0$ and therefore,
\[
 \tpl{x}^\top H\tpl{x} = \sum_{k=1}^C a_k\,\big(\tpl{v}_k^\top \tpl{x}\big)^2 = 0.
\]
Because $a_k > 0$, this implies that
\[
 \tpl{v}_k^\top \tpl{x} = 0 \qquad \text{for $k=1,2,\ldots,C$}.
\]
Because the set of vectors $\tpl{v}_k$ span $\R^C$,
\[
 \tpl{y}^\top \tpl{x} = 0 \qquad \text{for $\tpl{y}\in\R^C$}.
\]
This implies that $\tpl{x} = \tpl{0}$, but this contradicts our assumption.  We conclude that $H$ is not singular, and thus positive-definite.

From (\ref{eq:matrixG}) we construct the square root $\sqrt{G}$,
\[
 \sqrt{G} = \begin{pmatrix}
 \sqrt{\widetilde p_{i_1}} & 0 & \cdots & 0 \\[4pt]
 0 & \sqrt{\widetilde p_{i_2}} & \cdots & 0 \\[-2pt]
 \vdots & \vdots & & \vdots \\[2pt]
 0 & 0 & \cdots & \sqrt{\widetilde p_{i_C}}
 \end{pmatrix}.
\]
Multiplying the matrix $G H$ on the left by $\sqrt{G}^{-1}$ and on the right by $\sqrt{G}$, we get
\begin{equation}
 \sqrt{G}^{-1}  GH\, \sqrt{G} = \sqrt{G} \,H \sqrt{G}.
 \label{eq:similarity}
\end{equation}
Because the matrix $H$ is positive-definite, the matrix $\sqrt{G} \,H \sqrt{G}$ is also positive-definite, and therefore has positive real eigenvalues.  Equation~(\ref{eq:similarity}) expresses that the matrices $G H$ and  $\sqrt{G} \,H \sqrt{G}$ are similar, and therefore have the same eigenvalues.  As a result, all eigenvalues of the matrix $G H$ are positive and real.

To show that all eigenvalues of the matrix $G H$ are inside the unit circle, we first note that the matrix $H$ has positive components.  We construct an auxiliary matrix $H_+$ with components larger than $H$,
\begin{align}
 H_+
 &= \sum_{k=1}^{C-1} (h_{i_k}-h_{i_{k+1}})
 \; \frac{L\Big( \sum_{j=1}^k f_{i_j}\widetilde p_{i_j} \Big)}
 {\sum_{j=1}^k f_{i_j}\widetilde p_{i_j}}
 \; \tpl{v}_k \tpl{v}_k^\top + \; h_{i_C}
 \, \frac{L\Big( \sum_{j=1}^C f_{i_j}\widetilde p_{i_j} \Big)}
 {\sum_{j=1}^C f_{i_j}\widetilde p_{i_j}}
 \; \tpl{v}_C \tpl{v}_C^\top.
 \label{eq:matrixH*}
\end{align}
Comparing (\ref{eq:matrixH}) and (\ref{eq:matrixH*}) and using the inequality,
\[
 |L'(x)| = - L'(x) = \frac{1-\e^{-x}}{x^2} - \frac{x\,\e^{-x}}{x^2}
 \leq \frac{L(x)}{x} \qquad \text{for $x>0$},
\]
we see that
\[
 H \leq H_+ \qquad \text{(component-wise inequality)}.
\]
Because the matrix $G$ is diagonal with positive components, both $G H$ and $G H_+$ are matrices with positive components, and
\[
 G H \leq G H_+ \qquad \text{(component-wise inequality)}.
\]
Hence, the spectral radius of $G H$ is smaller than or equal to the spectral radius of $G H_+$.

We show that the vector $\tpl{w}$ with positive components,
\[
 \tpl{w} = \sum_{m=1}^C \widetilde p_{i_m} \tpl{e}_m,
\]
is an eigenvector with eigenvalue one of the matrix $G H_+$:
\begin{align*}
 G H_+ \tpl{w}
 &= \sum_{k=1}^{C-1} (h_{i_k}-h_{i_{k+1}})
 \; \frac{L\Big( \sum_{j=1}^k f_{i_j}\widetilde p_{i_j} \Big)}
 {\sum_{j=1}^k f_{i_j}\widetilde p_{i_j}}
 \; G \tpl{v}_k \tpl{v}_k^\top \tpl{w} + \; h_{i_C}
 \, \frac{L\Big( \sum_{j=1}^C f_{i_j}\widetilde p_{i_j} \Big)}
 {\sum_{j=1}^C f_{i_j}\widetilde p_{i_j}}
 \; G \tpl{v}_C \tpl{v}_C^\top \tpl{w} \\
 &= \sum_{k=1}^{C-1} (h_{i_k}-h_{i_{k+1}})
 \; L\Big( {\ds \sum_{j=1}^k f_{i_j}\widetilde p_{i_j}} \Big)
 \; G \tpl{v}_k + \; h_{i_C}
 \, L\Big( {\ds \sum_{j=1}^C f_{i_j}\widetilde p_{i_j}} \Big)
 \; G \tpl{v}_C \\
 &= \sum_{\ell=1}^C \Bigg( \sum_{k=\ell}^{C-1} (h_{i_k}-h_{i_{k+1}})
 \; L\Big( {\ds \sum_{j=1}^k f_{i_j}\widetilde p_{i_j}} \Big) + h_{i_C}
 \, L\Big( {\ds \sum_{j=1}^C f_{i_j}\widetilde p_{i_j}} \Big)
 \Bigg) f_{i_\ell} \widetilde p_{i_\ell} \tpl{e}_\ell \\
 &= \sum_{\ell=1}^C F_{i_\ell}(\tpl{\widetilde p})\,
 f_{i_\ell} \widetilde p_{i_\ell} \tpl{e}_\ell \\
 &= \sum_{\ell=1}^C \widetilde p_{i_\ell} \tpl{e}_\ell \\
 &= \tpl{w},
\end{align*}
where we have used (\ref{eq:equicondreda}) with $K=L$ and $i=i_\ell$ in the fourth equality and (\ref{eq:equicond}) in the fifth equality.  The matrix $G H_+$ has positive exponents, so we can apply Perron-Frobenius theory.  Because the vector $\tpl{w}$ has positive components and is an eigenvector of $G H_+$, we know that the corresponding eigenvalue equals the spectral radius of $G H_+$.  Hence, the spectral radius of $G H_+$ is equal to one, and the spectral radius of $G H$ is smaller than or equal to one.  We have shown before that the eigenvalues of $G H$ are real and positive.  We conclude that the eigenvalues of $J_\sC = \idty - G H$ belong to the real interval $[0,1)$.  This finishes the proof that conditions (\ref{eq:condstable}) are necessary and sufficient conditions for local stability of the equilibrium $\tpl{\widetilde p}$.

\section*{Appendix D: Proof of Result 3} 

Consider a set $\sS$ of species with traits $(g_i,h_i)$ satisfying Assumption~\ref{ass:generic}.  We prove that there is one and only one subset $\sC \subset \sS$ for which inequalities~(\ref{eq:condexistsome}) and (\ref{eq:condstable}) are simultaneously satisfied.

In the main text we have given a graphical representation of the coexistence conditions~(\ref{eq:condexistsome}) and the stability conditions~(\ref{eq:condstable}).  For a subset $\sC \subset \sS$ we construct a broken line passing through the trait pairs $(g_i,h_i)$ of species $i \in \sC$.  Conditions~(\ref{eq:condexistsome}) impose that at each species in $\sC$ the slope of the broken line should decrease (note that the slope changes at a species in $\sC$ due to Assumption~\ref{ass:generic}).  Conditions~(\ref{eq:condstable}) impose that each species in $\sS \setminus \sC$ should lie to the right of and below the broken line (note that no species in $\sS \setminus \sC$ lies on the broken line due to Assumption~\ref{ass:generic}).

\subsubsection*{Existence}

We present an explicit algorithm to construct a set $\sC$ satisfying conditions~(\ref{eq:condexistsome}) and (\ref{eq:condstable}):
\begin{center}
\hspace{0.1cm}
\begin{minipage}{.9\textwidth}
\begin{tabular}{p{0.2in}p{.8\textwidth}}
 1 & $\sC \gets \emptyset$ \\[6pt]
 2 & $\mathcal{A} \gets \ds \big\{ i\in\sS \,\big|\, h_i>g_i \big\}$ \\[6pt]
 3 & \textbf{if} $\mathcal{A} \neq \emptyset$ \textbf{then} \\[6pt]
 4 & $\qquad$ $i^* \gets \arg\max\limits_i
   \big\{\frac{h_i}{g_i}\,\big|\;i\in\mathcal{A}\big\}$ \\[6pt]
 5 & $\qquad$ $\sC \gets \sC\cup\{i^*\}$ \\[6pt]
 6 & $\qquad$ $\mathcal{A} \gets \ds \big\{ i\in\sS \,\big|\,
     g_i>g_{i^*},\; h_i-h_{i^*}>g_i-g_{i^*} \big\}$ \\[6pt]
 7 & $\qquad$ \textbf{while} $\mathcal{A} \neq \emptyset$ \bf{do} \\[6pt]
 8 & $\qquad$ $\qquad$ $i^* \gets \arg\max\limits_i
   \big\{\frac{h_i-h_{i^*}}{g_i-g_{i^*}}\,\big|\;i\in\mathcal{A}\big\}$ \\[6pt]
 9 & $\qquad$ $\qquad$ $\sC \gets \sC\cup\{i^*\}$ \\[6pt]
 10 & $\qquad$ $\qquad$ $\mathcal{A} \gets \ds \big\{ i\in\sS \,\big|\,
     g_i>g_{i^*},\; h_i-h_{i^*}>g_i-g_{i^*} \big\}$ \\[6pt]
 11 & $\qquad$ \textbf{end while} \\[6pt]
 12 & \textbf{end if} \\[6pt]
 13 & \textbf{return} $\sC$
\end{tabular}
\end{minipage}
\end{center}

Species are added one by one to the set $\sC$ (line 5 and 9).  Which species $i^*$ is added is determined by a maximization problem (line 4 and 8) over a set of species $\mathcal{A}$ (computed in line 2, 6 and 10).  Species are added in order of increasing $g$ (and increasing $h$).  Note that each maximization problem has a unique solution due to Assumption~\ref{ass:generic}.  The algorithm is illustrated in Figure~\ref{fig:intro} for an example of $S=5$ species.

The slope of the broken line decreases at each species $i$ in the set $\sC$ constructed in the algorithm.  Indeed, an increasing slope of the broken line at species $i$ would contradict that species $i$ was obtained by maximizing the slope.  Hence, conditions~(\ref{eq:condexistsome}) are satisfied.  Similarly, each species $j$ not belonging to the set $\sC$ lies to the right of and below the broken line corresponding to the set $\sC$.  Indeed, if species $j$ would lie to the left of and above the broken line, then there would be at least one maximization problem for which the line segment to species $j$ has a larger slope than the maximal slope.  From this contradiction we conclude that conditions~(\ref{eq:condstable}) are satisfied for the set $\sC$ constructed in the algorithm.

\subsubsection*{Uniqueness}

We show that there is only one set $\sC$ for which conditions~(\ref{eq:condexistsome}) and (\ref{eq:condstable}) are satisfied.

Suppose that there are two sets $\sC_a$ and $\sC_b$ for which conditions~(\ref{eq:condexistsome}) and (\ref{eq:condstable}) hold.  Assume $\sC_a$ and $\sC_b$ have $C_a$ and $C_b$ elements, respectively,
\begin{align*}
 \sC_a = \{ i_1,i_2,\ldots,i_{C_a} \}
 \qquad & \text{with} \quad g_{i_1} < g_{i_2} < \ldots < g_{i_{C_a}} \\
 \sC_b = \{ j_1,j_2,\ldots,j_{C_b} \}
 \qquad & \text{with} \quad g_{j_1} < g_{j_2} < \ldots < g_{j_{C_b}}.
\end{align*}
We assume without loss of generality that $C_a \leq C_b$.

We consider the broken lines corresponding to $\sC_a$ and $\sC_b$.  Both broken lines start at the origin $0$.  We compare the first species~$i_1$ and $j_1$ of the broken lines.  Suppose $g_{i_1} > g_{j_1}$.  Species~$i_1$ cannot lie above the broken line of $\sC_b$ (because (\ref{eq:condstable}) holds for $\sC_b$).  Hence, species~$i_1$ lies on or below the broken line of $\sC_b$.  But then species $j_1$ lies above the segment $[0,i_1]$ which is not possible (because (\ref{eq:condstable}) holds for $\sC_a$).  Hence, $g_{i_1} \leq g_{j_1}$.  By symmetry we also have $g_{i_1} \geq g_{j_1}$, so that $g_{i_1} = g_{j_1}$.  By Assumption~\ref{ass:generic} this implies that $i_1 = j_1$.  Next, we compare the second species~$i_2$ and $j_2$ of the broken lines.  A similar argument leads to $i_2 = j_2$.  By repeating the same argument, we obtain $i_3 = j_3$, \ldots, $i_{C_a} = j_{C_a}$.  Left of species $i_{C_a} = j_{C_a}$ the broken line of $\sC_a$ is a half-line with slope one.  Species $j_{C_a+1}$ cannot lie above the half-line (because (\ref{eq:condstable}) holds for $\sC_a$).  Species $j_{C_a+1}$ cannot lie below the half-line because then the segment $[j_{C_a},j_{C_a+1}]$ would have slope smaller than one.  Assumption~\ref{ass:generic} excludes that $j_{C_a+1}$ lies on the half-line.  Hence, the broken line of $\sC_b$ left of $i_{C_a} = j_{C_a}$ is also a half-line with slope one.  We conclude that $C_a = C_b$ and that the sets $\sC_a$ and $\sC_b$ are identical.

Finally, we note that alternative (but equivalent) constructions of the set $\sC$ exist.  We mention a construction in terms of the convex hull.  For each species $i$ let $D_i$ be the half-line $h=h_i+g-g_i$, $g\geq g_i$.  Let $D$ be the convex hull of the half-line $h=0$, $g\geq 0$ and the half-lines $D_i$.  The set $\sC$ satisfying conditions~(\ref{eq:condexistsome}) and (\ref{eq:condstable}) is the the set of species lying on the boundary of $D$.  The species not belonging to $\sC$ lie in the interior of $D$.  This construction can also be used to prove existence and uniqueness of the set $\sC$ of coexisting species.

\section*{Appendix E: Model with overlapping generations} 

We demonstrate how our analysis for the discrete-time system~(\ref{eq:discgen}) can be extended to the continuous-time system~(\ref{eq:contgen}).  We consider the conditions for the existence of an equilibrium with a specific coexistence set, and the corresponding conditions for local stability.

Dynamical system~(\ref{eq:contgen}) can be written as
\[
 \frac{\diff p_i}{\diff t} = m \Big( F_i(\tpl{p})\, f_i p_i - p_i \Big).
\]
with $m$ the mortality rate and $F_i(\tpl{p})$ defined in (\ref{eq:defF}).  The equilibrium conditions, obtained by setting the right-hand side of the latter equations equal to zero, are identical to the equilibrium conditions (\ref{eq:equicond}) for the discrete-time system.  Hence, the conditions (\ref{eq:condexistsome}) for the existence of an equilibrium with a specific coexistence set and the explicit expressions (\ref{eq:equilib}) for the equilibrium occupancies also hold for the continuous-time system.

Concerning the stability conditions, note that the Jacobian matrix $J_\text{cont}$ (evaluated at an equilibrium) of the continuous-time system~(\ref{eq:contgen}) is related to the Jacobian matrix $J$ (evaluated at an equilibrium) of the discrete-time system~(\ref{eq:discgen}) by
\[
 J_\text{cont} = m \big( J - \idty \big).
\]
We have shown that conditions~(\ref{eq:condstable}) guarantee that all eigenvalues of $J$ belong to the real interval $[0,1)$.  This implies that the eigenvalues of $J_\text{cont}$ belong to the real interval $[-m,0)$, so that the equilibrium of the continuous-time system is locally stable.  Conversely, we have shown that if one of conditions~(\ref{eq:condstable}) is violated, then there exists an eigenvalue $\lambda>1$ of $J$ (assuming genericity of parameters).  This implies that there exists an eigenvalue $m(\lambda-1)>0$ of $J_\text{cont}$, so that the equilibrium of the continuous-time system is unstable.  This proves that conditions~(\ref{eq:condstable}) are necessary and sufficient for local stability of an equilibrium of the continuous-time system as well.

\subsubsection*{Note on \citet{ml10}}

\citet{ml10} conjectured that the following conditions are necessary and sufficient for the existence of an equilibrium in which all $S$ species coexist:
\begin{subequations}\label{eq:muller1}
\begin{align}
 h_1-h_2 &> \frac{f_2-f_1}{f_1f_2} \label{eq:muller1a} \\
 \frac{f_ih_i-f_{i-1}h_{i-1}}{f_i-f_{i-1}}
 &> \frac{f_{i+1}h_{i+1}-f_ih_i}{f_{i+1}-f_i} + \frac{1}{f_i}
 \qquad \text{for all}\ i=2,3,\ldots,S-1 \label{eq:muller1b} \\
 f_Sh_S &> f_{S-1}h_{S-1} + \frac{f_S-f_{S-1}}{f_S}. \label{eq:muller1c}
\end{align}
\end{subequations}
We compare conditions~(\ref{eq:muller1}) with the exact necessary and sufficient conditions, which follow from Result~\ref{res:equilib},
\begin{subequations}\label{eq:muller2}
\begin{align}
 \frac{h_1-h_2}{g_1-g_2} &> 1 \label{eq:muller2a} \\
 \frac{h_i-h_{i+1}}{g_i-g_{i+1}} &> \frac{h_{i-1}-h_i}{g_{i-1}-g_i}
 \qquad \text{for all}\ i=2,3,\ldots,S-1 \label{eq:muller2b} \\
 \frac{h_S}{g_S} &> \frac{h_{S-1}-h_S}{g_{S-1}-g_S}. \label{eq:muller2c}
\end{align}
\end{subequations}
Inequality~(\ref{eq:muller1a}) is equivalent with inequality~(\ref{eq:muller2a}).  Inequality (\ref{eq:muller1b}) with the last term in the right-hand side dropped is equivalent with inequalitiy~(\ref{eq:muller2b}).  Inequality~(\ref{eq:muller1c}) with the last term dropped is equivalent with inequality~(\ref{eq:muller2c}).  Hence, conditions~(\ref{eq:muller1}) are too strong.

\end{document}